\documentclass[aps,prx,twocolumn,superscriptaddress,showpacs,floatfix]{revtex4-1}
\usepackage{amsmath,amssymb,graphicx}

\usepackage[utf8]{inputenc}
\usepackage[T1]{fontenc}
\usepackage{xcolor}

\IfFileExists{newtxtext.sty}
   {\usepackage{newtxtext,newtxmath}}
   {\IfFileExists{stix.sty}
      {\usepackage{stix}}
      {\IfFileExists{mathptmx.sty}
      {\usepackage{mathptmx}}{} } }

\usepackage{textcomp}

\usepackage{bm}

\IfFileExists{siunitx.sty}{\usepackage{booktabs,siunitx}}{}

\pdfoutput=1
\usepackage{color}
\definecolor{LinkColor}{rgb}{0.256,0.439,0.588}
\usepackage{hyperref}
\hypersetup{
   pdfauthor={Zi-Hong Liu, Xiao Yan Xu, Yang Qi, and Zi Yang Meng},
   pdftitle={Functional quantum Monte Carlo Method},
   colorlinks=true,
   citecolor=LinkColor,
   linkcolor=LinkColor,
   urlcolor=LinkColor
}

\renewcommand{\vec}[1]{\mathbf{#1}}

\usepackage{pifont}

\begin{document}

\title{Itinerant Quantum Critical Point with Fermion Pockets and Hot Spots}

\author{Zi Hong Liu}
\affiliation{Beijing National Laboratory for Condensed Matter Physics and Institute of Physics, Chinese Academy of Sciences, Beijing 100190, China}
\affiliation{School of Physical Sciences, University of Chinese Academy of Sciences, Beijing 100190, China}

\author{Gaopei Pan}
\affiliation{Beijing National Laboratory for Condensed Matter Physics and Institute of Physics, Chinese Academy of Sciences, Beijing 100190, China}
\affiliation{School of Physical Sciences, University of Chinese Academy of Sciences, Beijing 100190, China}

\author{Xiao Yan Xu}
\affiliation{Department of Physics, Hong Kong University of Science and Technology, Clear Water Bay, Hong Kong, China}

\author{Kai Sun}
\affiliation{Department of Physics, University of Michigan, Ann Arbor, MI 48109, USA}

\author{Zi Yang Meng}
\affiliation{Beijing National Laboratory for Condensed Matter Physics and Institute of Physics, Chinese Academy of Sciences, Beijing 100190, China}
\affiliation{Department of Physics, The University of Hong Kong, Pokfulam Road, Hong Kong, China}
\affiliation{CAS Center of Excellence in Topological Quantum Computation and School of Physical Sciences, University of Chinese Academy of Sciences, Beijing 100190, China}
\affiliation{Songshan Lake Materials Laboratory, Dongguan, Guangdong 523808, China}

\begin{abstract}
	Metallic quantum criticality is among the central theme in the understanding of correlated electronic systems, and converging results between analytical and numerical approaches are still under calling. In this work, we develop state-of-art large scale quantum Monte Carlo simulation technique and systematically investigate the itinerant quantum critical point on a 2D square lattice with antiferromagnetic spin fluctuations at wavevector $\mathbf{Q}=(\pi,\pi)$ -- a problem that resembles the Fermi surface setup and low-energy antiferromagnetic fluctuations in high-Tc cuprates and other critical metals, which might be relevant to their non-Fermi-liquid behaviors. System sizes of $60\times 60 \times 320$ ($L \times L \times L_\tau$) are comfortably accessed, and the quantum critical scaling behaviors are revealed with unprecedingly high precision. We found that  the antiferromagnetic spin fluctuations introduce effective interactions among fermions and the fermions in return render the bare bosonic critical point into a new universality, different from both the bare Ising universality class and the Hertz-Mills-Moriya RPA prediction. At the quantum critical point, a finite anomalous dimension $\eta\sim 0.125$ is observed in the bosonic propagator, and fermions at hot spots evolve into a non-Fermi-liquid. In the antiferromagnetically ordered metallic phase, fermion pockets are observed as energy gap opens up at the hot spots. These results bridge the recent theoretical and numerical developments in metallic quantum criticality and can be served as the stepping stone towards final understanding of the 2D correlated fermions interacting with gapless critical excitations.
\end{abstract}

\date{\today}

\maketitle

\section{Introduction}
\label{sec:intro}
In the study of correlated materials, quantum criticality in itinerant electron systems is of great importance and interests~\cite{Hertz1976,Millis1993,Moriya1985,Stewart2001,Chubukov2004,Belitz2005,Loehneysen2007,Chubukov2009}. It plays a vital role in the understanding of anomalous transport, strange metal and non-fermi-liquid behaviors~\cite{Metzner2003,Senthil2008,Holder2015,Metlitski2015,Xu2017} in heavy-fermion materials~\cite{Custers2003,Steppke2013}, Cu- and Fe-based high-temperature superconductors~\cite{ZhangWenLiang2016,LiuZhaoYu2016,Gu2017} as well as the recently discovered  pressure-driven quantum critical point (QCP) between magnetic order and superconductivity in transition-metal monopnictides, CrAs~\cite{Wu2014}, MnP~\cite{Cheng2015}, CrAs$_{1-x}$P$_x$~\cite{JGCheng2018} and other Cr/Mn-3d electron systems~\cite{Cheng2017}. However, despite of extensive efforts in decades~\cite{Hertz1976,Millis1993,Moriya1985,Stewart2001,Metzner2003,Abanov2003,Abanov2004,Chubukov2004,Belitz2005,Loehneysen2007,Chubukov2009,Metlitski2010a,Metlitski2010b,Sur2016,Schlief2017,SSLee2018,Schlief2018}, itinerant quantum criticality is still among the most challenging subjects in condensed matter physics, due to its nonperturbative nature, and many important questions and puzzles remain open.

The recent development in sign-problem-free quantum Monte Carlo techniques has paved a new path way towards sharpening our understanding about this challenging problem (see a concise commentary in Ref.~\cite{Chubukov2018JC} and a review in Ref.~\cite{XiaoYanXu2019_review} that summarize the recent progress). Via coupling a Fermi liquid with various bosonic critical fluctuations,  a wide variety of itinerant quantum critical systems have been studied, such as Ising-nematic~\cite{Schattner2015a,Lederer2016}, ferromagnetic~\cite{Xu2017}, charge density wave~\cite{ZXLi2015}, spin density wave~\cite{Berg12,ZXLi2016,Schattner2015b,Gerlach2017,ZHLiu2017,ZHLiu2018} and interaction-driven topological phase transitions and gauge fields~\cite{Xu2016a,Assaad2016,Gazit2016,He2018,XiaoYanXu2019_PRX,ChuangChen2019OM}. With the fast development in QMC techniques, in particular the self-learning Monte Calro (SLMC)~\cite{liu2016self,liu2016fermion,Xu2016self,Nagai2017,HTShen2018,ChenChuang2018,ChenChuang2018Dirac} and elective momentum ultra-size quantum Monte Carlo (EQMC)~\cite{ZHLiu2018}, it now becomes possible to explore larger system sizes than those handled with conventional determinantal quantum Monte Carlo, and consequently allowing us to access the genuine scaling behaviors in the infrared (IR) limit for itinerant quantum criticality. 

Although a lot intriguing results and insights have been obtained, for the search of novel quantum critical points beyond the Hertz-Millis-Moriya theory, a major gap between theory and numerical studies still remain. So far, in QMC simulations, among all recently studied itinerant QCPs, either the Hertz-Millis mean-field scaling behavior is found~\cite{Schattner2015a,ZHLiu2017}, or unpredicted exponents, deviating from existing theories, are observed~\cite{Xu2017}, while theoretically proposed properties beyond the Hertz-Millis-Moriya scaling behaviors still remain to be numerically observed and verified. 

In this paper, we aim at improving the convergence between theoretical and numerical studies by focusing on itinerant QCPs with finite ordering wave vector $\mathbf{Q}\ne 0$, e.g. charge/spin density waves (CDW/SDW). One key question in the study of these QCPs is about the universality class, i.e. whether all these types of QCPs, e.g. commensurate and incommensurate CDW/SDW QCPs, belong to the same universality class or not. To the leading order, within the random phase approximation (RPA), as long as the ordering wave vector $\mathbf{Q}$ is smaller than twice the Fermi wavevector $2 k_F$, or more precisely, as we shift the Fermi surface (FS) by the ordering wave vector  $\mathbf{Q}$ in the momentum space, the shifted FS and the original one shall cross at hot spots, instead of tangentially touching with or overrun each other, the same (linear $\omega$) Landau damping and critical dynamics is predicated regardless of microscopic details, implying the dynamic critical exponent $z=2$. For a 2D system, this makes the effective dimensions $d+z=4$, coinciding with the upper critical dimension. As a result, within the Hertz-Millis approximation, mean-field critical exponents shall always be expected, up to possible logarithmic corrections, and thus all these QCPs belong to the same universality class~\cite{Hertz1976,Millis1993,Moriya1985,KaiSun2008}. 

On the other hand, more recent theoretical developments point out that this conclusion becomes questionable once higher order effects are taken into account. In particular, two different universality classes need to be distinguished, depending on whether $2 \mathbf{Q}$ coincides with a lattice vector or not, which will be dubbed as $2\mathbf{Q}=\Gamma$ and $2\mathbf{Q}\neq \Gamma$ to demonstrate that $2 \mathbf{Q}$, mod a reciprocal lattice vector, coincides or not with the $\Gamma$ point. Among these two cases, $2\mathbf{Q}=\Gamma$ (e.g. antiferromagnetic QCP with $\mathbf{Q}=(\pi,\pi)$) is highly exotic. As Abanov and coworkers pointed out explictly, in this case the Hertz-Millis mean-field scaling will breaks down and a nonzero anomalous dimension shall emerge~\cite{Abanov2003}.  In addition, the critical fluctuations will also change the fermion dispersion near the hot spots, resulting in a critical-fluctuation-induced Fermi surface nesting: i.e. even if one starts from a Fermi surface without nesting, the RG (renormalization group) flow of the Fermi velocity will deform the Fermi surface at hotspots towards nesting~\cite{Abanov2003}. This Fermi surface deformation will further increases the anomalous dimension and make the scaling exponent deviate even further from Hertz-Millis prediction~\cite{Abanov2003,Abanov2004}, and even modifies the dynamic critical exponent $z$, as pointed out explicitly by Metlitski and Sachdev and others~\cite{Metlitski2010b,Schlief2017,SSLee2018}.  For $2\mathbf{Q} \ne \Gamma$, on the other hand, these exotic behaviors are not expected, at least up to the same order in the $1/N$ expansion and thus presumably, they follows the Hertz-Millis mean-field scaling relation.

On the numerical side, a QCP with $2 \mathbf{Q}\ne \Gamma$ was recently studied~\cite{ZHLiu2017,ZHLiu2018} and the results are in good agreement with the Hertz-Millis-Moriya theory. For the more exotic case with $2 \mathbf{Q}=\Gamma$, the numerical result is less clear because in QMC simulations,  a superconducting dome usually arises and covers the QCP~\cite{Schattner2015b,Gerlach2017,Berg2018}. Outside the superconducting dome, at some distance away from the QCP, mean-field exponents are observed to be consistent with the Hertz-Millis-Moriya theory. However, whether the predicted anomalous (non-mean-field) behaviors~\cite{Abanov2003,Metlitski2010b,Schlief2017,SSLee2018} will arise in the close vicinity of the QCP remains an open question, which requires the suppression of the superconducting order. In addition, due to the divergent length scale at a QCP, to obtain reliable scaling exponents, large system sizes is necessary to overcome the finite-size effect.

In this manuscript, we perform large-scale quantum Monte Carlo simulations to study the antiferromagnetic metallic quantum critical point (AFM-QCP) with $2 \mathbf{Q}=\Gamma$. In this study, two main efforts are made in order to accurately obtain the critical behavior in the close vicinity of the QCP. (1) We design a lattice model that realizes the desired AFM-QCP with superconducting dome greatly suppressed to expose the quantum critical regions and (2) we employ the determinantal quantum Monte Carlo (DQMC) as well as the elective momentum ultra-size QMC (EQMC), both with self-learning updates to access much larger systems sizes beyond existing efforts. The more conventional DQMC technique allow us to access system sizes up to $28\times 28 \times200$ for $L\times L\times L_{\tau}$ for 2D square lattice, while EQMC can access much larger sizes ($60\times 60 \times 320$) to further reduce the finites-size effect and confirms scaling exponents with higher accuracy.
These two efforts (1) and (2) allow us to access the metallic quantum critical region and to reveal its IR scaling behaviors with great precision, where we found a large anomalous dimensions significantly different from the Hertz-Millis theory prediction, and we also observed that the Fermi surface near the hotspots rotates towards nesting at the QCP, as predicted in the RG analysis~\cite{Abanov2003,Metlitski2010b,Schlief2017,SSLee2018}. And quantitative comparison between theory and numerical results is also performed. These results bridge the recent theoretical and numerical developments and are the precious stepping stone towards final understanding of the metallic quantum criticality in 2D. 

\begin{figure*}[htp!]
	\centering
	\includegraphics[width=\textwidth]{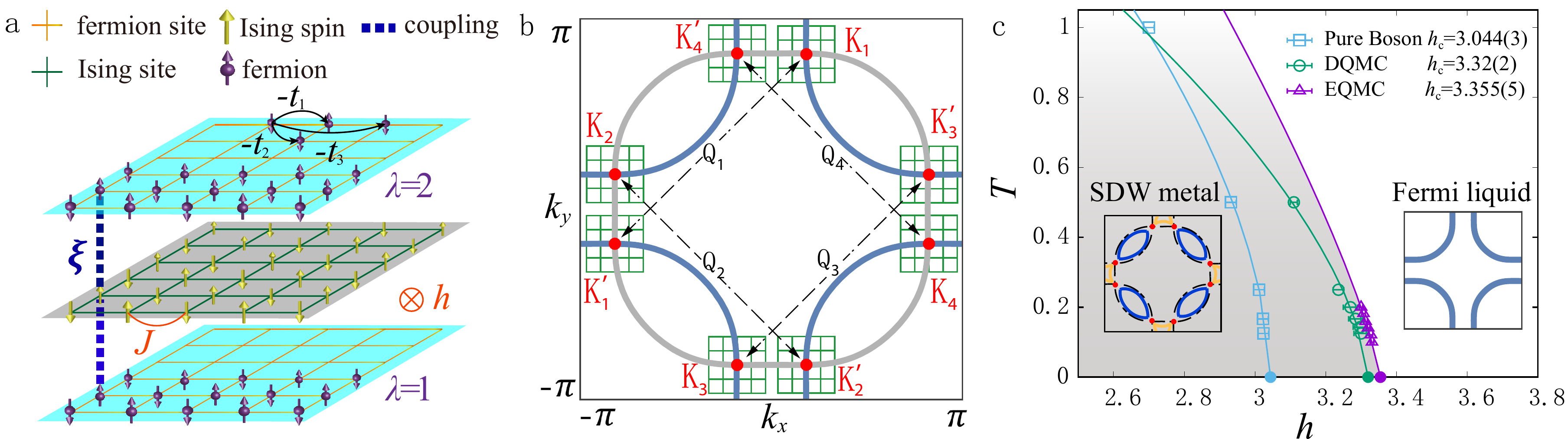}
	\caption{(a) Illustration of the model in Eq.~\eqref{eq:HfM}. Fermions reside on two of the layers ($\lambda$ = 1,2) with intra-layer nearest, 2nd and 3rd neighbor hoppings $t_1$, $t_2$ and $t_3$. The middle layer is composed of Ising spins $s^{z}_{i}$, subject to nearest-neighbor antiferromagnetic Ising coupling $J$ and a transverse magnetic field $h$. Between the layers, an onsite Ising coupling is introduced between fermion and Ising spins ($\xi$). (b) Brilliouin zone (BZ) of the model in Eq.~\eqref{eq:HfM}. The blue lines are the fermi surface (FS) of $H_{f}$ and $\mathbf{Q}_i=(\pm \pi, \pm \pi), \; i=1,2,3,4$ are the AFM wavevectors, and the four pairs of  $\{\mathbf{K}_i,\mathbf{K'}_i\}, \; i=1,2,3,4$ are the position of the hot spots (red dots), each pair is connected by a $\mathbf{Q}_i$ vector. The folded FS (gray lines), coming from translating the bare FS by momentum $\mathbf{Q}_i$. The green patches show the $\mathbf{k}$ mesh built around hot spots, number of momentum points inside each patch is denoted as $N_f$. (c) Phase diagram of model Eq.~\eqref{eq:HfM}. The light blue line marks the phase boundaries of the pure bosonic model $H_{b}$, with a QCP (light blue dot) at $h_c=3.044(3)$~\cite{Bloete2002,Hesselmann2016} with 3D Ising universality. After coupling with fermions, the QCP shifts to higher values. The green solid dot is the QCP obtained with DQMC ($h_c=3.32(2)$). The violet solid dot is the QCP obtained from EQMC ($h_c=3.355(5)$), although the position of the QCP shifts, as it is an nonuniversal quantity, the scaling behavior inside the quantum critical region is consistent between DQMC and EQMC. The EQMC scheme can comfortably capture the IR physics of AFM-QCP, with much larger system sizes, $60\times 60\times 320$, compared with that in DQMC with $28\times28\times 200$. The procedure of how the phase boundary is determined is shown in the Sec. I in the Appendix.}
	\label{fig:fig1}
\end{figure*}


\section{Model and Method}
\label{sec:method}

\subsection{Antiferromagnetic fermiology}
\label{sec:ideas}

The square lattice AFM model that we designed are schematically shown in Fig.~\ref{fig:fig1}(a) with two fermion-layers and one Ising-spin-layer in between. Fermions are subject to intra-layer nearest, second and third neighbor hoppings $t_1$, $t_2$ and $t_3$, as well as the chemical potential $\mu$. The Ising spin layer is composed of Ising spins $s^{z}_{i}$ with nearest neighbor antiferromagnetic coupling $J$ ($J>0$) and a transverse magnetic field $h$ along $s^{x}$. Fermions and Ising spins are coupled together via an inter-layer onsite Ising coupling $\xi$. The Hamiltonian is given as
\begin{equation}
	\label{eq:HfM}
	H = H_{f} + H_{b} + H_{fb}
\end{equation}
where
\begin{align}
	H_{f}  = & -t_1\sum_{\left \langle ij \right \rangle, \lambda ,\sigma }c_{i,\lambda, \sigma}^\dagger c_{j,\lambda ,\sigma}-t_2\sum_{\left \langle \left \langle ij \right \rangle \right \rangle, \lambda ,\sigma }c_{i,\lambda ,\sigma}^\dagger c_{j,\lambda ,\sigma}\nonumber \\
	& -t_3\sum_{\left \langle \left \langle \left \langle ij \right \rangle \right \rangle \right \rangle ,\lambda ,\sigma }c_{i,\lambda ,\sigma}^\dagger c_{j,\lambda, \sigma}+h.c.-\mu\sum_{i,\lambda ,\sigma}n_{i,\lambda ,\sigma} \\
	H_{b}= & J\sum_{\left \langle ij \right \rangle }s_i^z s_j^z-h\sum_{i}s_i^x \\
	H_{fb}= & -\xi \sum_{i}s_i^z\left (\sigma_{i,1}^z + \sigma_{i,2}^z \right),
\end{align}
and $\sigma^{z}_{i,\lambda}=\frac{1}{2}(c^{\dagger}_{i,\lambda,\uparrow}c_{i,\lambda,\uparrow}-c^{\dagger}_{i,\lambda,\downarrow}c_{i,\lambda,\downarrow})$ is the fermion spin along $z$.

$H_{b}$ describes a 2D transverse-field Ising model and has a phase diagram spanned along the axes of temperature $T$ and $h$. As shown in Fig.~\ref{fig:fig1} (c) (light blue line), at $h=0$, the system undergoes a 2D Ising thermal transition at a finite $T$. Gradually turning on a finite $h$, the system experiences the same AFM 2D Ising transition with a lower transition temperature, until $h=h_c$ ($3.04438(2)$), where the transition turns into a $T=0$ quantum phase transition in the 3D Ising universality class~\cite{Bloete2002,Hesselmann2016}. Such an antiferromagnetic order has a wave vector $\mathbf{Q}=(\pi,\pi)$, as denoted by the $\mathbf{Q}_{i}$ with $i=1,2,3,4$ in Fig.~\ref{fig:fig1}(b). 

The fermions in $H_f$ experience the AFM fluctuations in $H_b$ via the fermion-spin coupling $H_{fb}$. As shown in the Fig.~\ref{fig:fig1} (b), the original FS and FSs from zone folding [shifted by the ordering wave vector $\mathbf{Q}_i$ ($i=1,2,3,4$)] form Fermi pockets and the so-called hot spots, which are crossing points between the original and the folded FSs labeled as $\mathbf{K}_i$ and  $\mathbf{K}_i'$ with $i=1,2,3,4$. 

In the simulation, we set $t_1=1.0$, $t_2=-0.32$, $t_3=0.128$, $J=1$, $\mu=-1.11856$ (electron density $\langle n_{i,\lambda}\rangle \sim 0.8$), the coupling strength $\xi=1.0$ and leave $h$ as control parameters. The parameters are chosen according to Ref.~\cite{Chowdhury2014}, such that deep in the AFM phase of Ising spins, the FS exhibits four big Fermi pockets and four pairs of hot spots (hot spot number $N_{h.s.}=8\times2=16$ where the factor $2$ comes from two fermion layers) as a result of band folding due to AFM ordering. Such an antiferromagnetic fermiology is summarized in Fig.~\ref{fig:fig1} (b).

\subsection{Ising scaling for the bare boson model}
\label{sec:boson}

\begin{figure}[htp]
	\centering
	\includegraphics[width=\columnwidth]{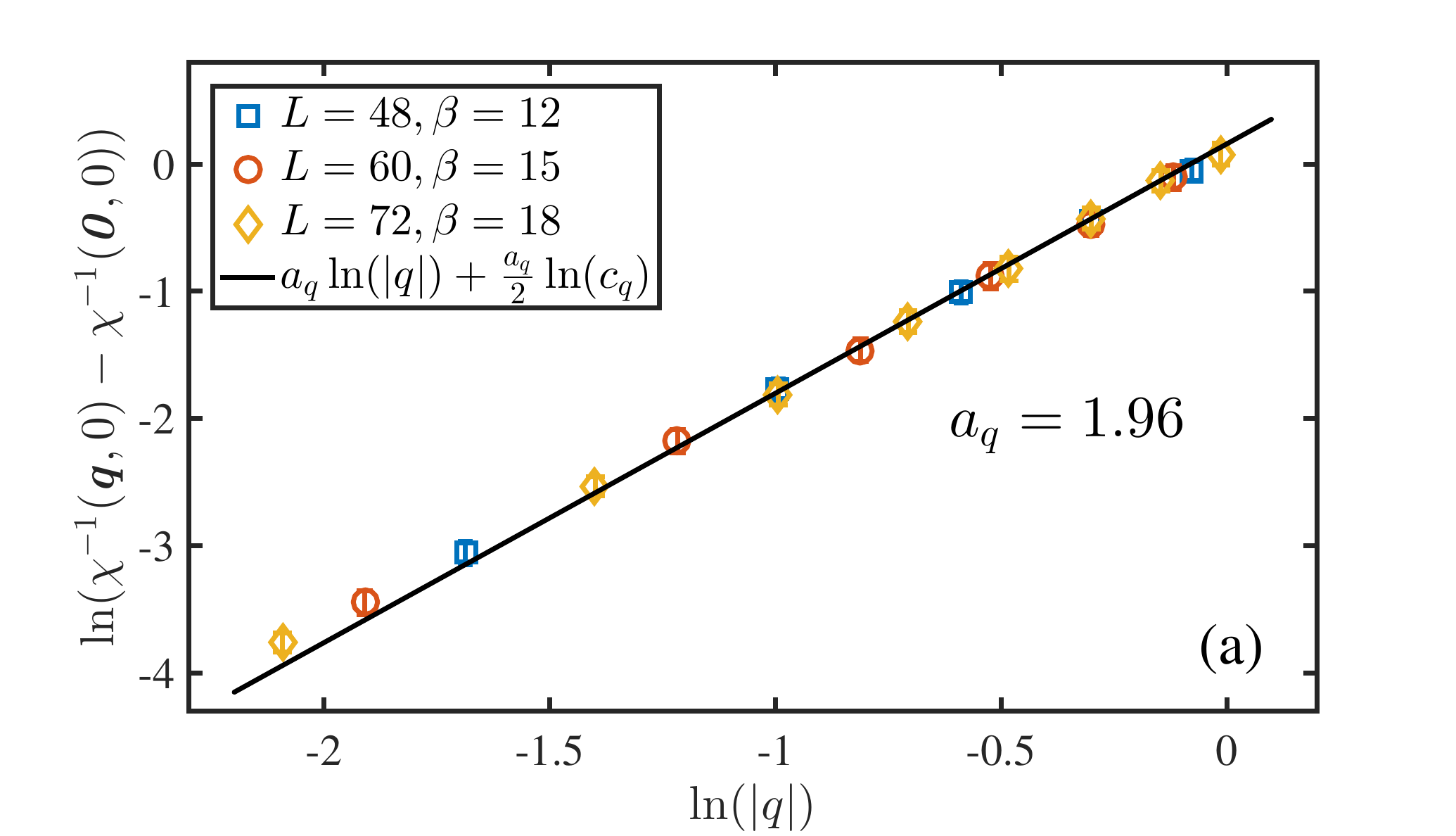}
	\includegraphics[width=\columnwidth]{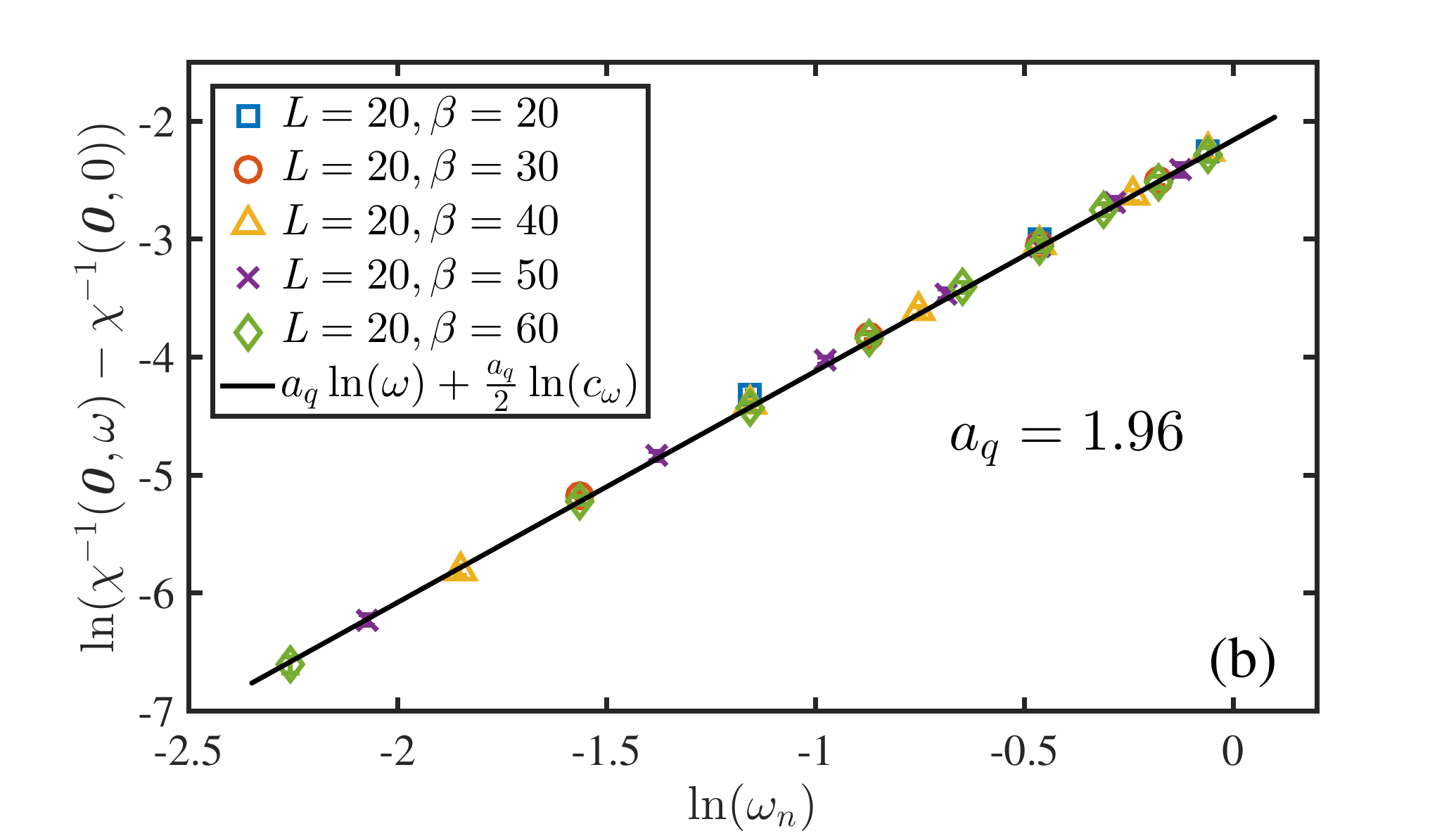}
	\caption{(a) Momentum dependence of the $\chi(\mathbf{q},\omega=0)$ at $h=h_c$, for the bare boson model $H_{b}$. The system sizes are $L=48,60,72$ respectively and to achieve quantum critical scaling, $\beta\propto L$ is applied. The line goes through the data points is $a_q\ln(|q|)+\frac{a_q}{2}\ln(c_q)$, with $a_q=2-\eta=1.96$.  (b) Frequency dependence of $\chi(\mathbf{q=0},\omega)$ at $h=h_c$, for the bare boson model $H_b$. The system sizes is $L=20$ with increasing $\beta=20,30,40,50$ and $60$. The line goes through the data points is  $a_q\ln(\omega)+\frac{a_q}{2}\ln(c_\omega)$, with $a_q=2-\eta=1.96$.}
	\label{fig:fig2}
\end{figure}

Before presenting our results about the itinerant AFM-QCP, we first discuss the QCP in the pure boson limit without fermions, which serves as a benchmark for the non-trivial itinerant quantum criticality. It is known that the pure boson QCP ($H_b$)  belongs to the $(2+1)$D Ising universality class~\cite{Bloete2002,Hesselmann2016,Xu2017}. This can be demonstrated numerically by calculating 
the dynamic spin susceptibility 
\begin{equation}
	\chi(T,h,\vec{q},\omega_{n})=\frac{1}{L^{2}}\sum_{ij}\int_{0}^{\beta}d\tau e^{i\omega_{n}\tau-i\vec{q}\cdot\vec{r}_{ij}}\langle s^{z}_{i}(\tau)s^{z}_{j}(0)\rangle.
	\label{eq:susceptibility}
\end{equation}
In principle, near the QCP, the functional form of the dynamic spin susceptibility is complicated and hard to write down explicitly. However, in the quantum critical region, as we set $h=h_c$, the scaling functional form of $\chi(T,h_c,\vec{q},\omega_{n})$ can be  described by the following asymptotic form 
\begin{equation}
	\chi(T,h_c,\mathbf{q},\omega_n)= \frac{1}{c_{t}T^{2}+(c_q |\mathbf{q}|^{2} + c_{\omega}\omega^{2})^{a_{q}/2}},
	\label{eq:susceptibility_omega_k}
\end{equation}
where $a_q=2-\eta=1.964(2)$ is the universal critical exponents of the 3D Ising universal class, and $c_t$, $c_q$ and $c_{\omega}$ are non-universal coefficients. This scaling form [Eq.~\eqref{eq:susceptibility_omega_k}] explicitly respects the emergent Lorentz symmetry at the Ising critical point.

Without fermions, we can employ the standard path-integral scheme to map the 2D transverse-field Ising model to a $(2+1)$D classical Ising model~\cite{YCWang2017}. To solve this anisotropic 3D Ising model with Monte Carlo simulations, we performed Wolff~\cite{Wolff1989} and Swendsen-Wang~\cite{Swendsen1987} cluster updates to access sufficiently large system sizes and low temperature. The dynamic susceptibility are shown in Fig.~\ref{fig:fig2}. To explore the momentum dependence of the susceptibility, we plot $\chi^{-1}(T,h_c,\mathbf{q},\omega=0)-\chi^{-1}(T,h_c,\mathbf{q}=0,\omega=0)$, where the momentum $\mathbf{q}$ is measured from $\mathbf{Q}=(\pi,\pi)$ and the substraction is to get rid of the finite temperature background, such that the following scaling relation is expected at low $T$
\begin{equation}
	\chi^{-1}(T,h_c,\mathbf{q},\omega=0)-\chi^{-1}(T, h_c ,\mathbf{q}=0,\omega=0) = c_q^{a_q/2}|\mathbf{q}|^{a_q}.
\end{equation}

As shown in Fig.~\ref{fig:fig2} (a) such a scaling relation is indeed observed with $L=48,60$ and $72$ with $\beta=1/T\propto L$. The power-law divergence of the $(2+1)$D Ising quantum critical susceptibility with power $a_q=2-\eta=1.96$ is clearly revealed (the 3D Ising anomalous dimension $\eta=0.04$.)

Similar scaling relation is also observed in the frequency dependence as shown in Fig.~\ref{fig:fig2} (b), where we plot
\begin{equation}
	\chi^{-1}(T, h_c,0,\omega)-\chi^{-1}(T,h_c,0,\omega=0) = c_\omega^{a_q/2} \omega^{a_q}
\end{equation}
for $L=20$ with increasing $\beta$. The expected power-law decay with the small anomalous dimension $a_q=2-\eta=1.96$ is clearly obtained. Hence, the data in Fig.~\ref{fig:fig2} (a) and (b) confirm the QCP for the pure boson part $H_{b}$ in Eq.~\eqref{eq:HfM} belongs to the $(2+1)$D Ising universality class.

\subsection{DQMC and EQMC}
\label{sec:QMC}

To solve the problem in Eq.~\eqref{eq:HfM} we employ two complementary fermionic quantum Monte Carlo schemes. 

The first one is the standard DQMC~\cite{BSS1981,Hirsch_1983,AssaadEvertz2008,Xu2017} with self-learning Monte Carlo update scheme (SLMC)~\cite{liu2016self,liu2016fermion,Xu2016self,Nagai2017,HTShen2018,ChenChuang2018,
	ChenChuang2018Dirac} to speedup the simulation. In SLMC, we first perform the standard DQMC simulation on the model in Eq.~\eqref{eq:HfM}, and then train an effective boson Hamiltonian that contains long-range two-body interactions both in spatial and temporal directions. The effective Hamiltonian serves as the proper low-energy description of the problem at hand with the fermion degree of freedom integrated out. We then use the effective Hamiltonian to guide the Monte Carlo simulations, i.e., we will perform many sweeps of the effective bosonic model (as the computational cost of updating the boson model is $O(\beta N)$, dramatically lower than the update of fermion determinant which scales as $O(\beta N^3)$, and then evaluate the fermion determinant of the original model in Eq.~\eqref{eq:HfM} such that the detailed balance
of the global update is satisfied. As shown in our previous works~\cite{liu2016fermion,Xu2016self,ChenChuang2018,ZHLiu2017,ChenChuang2018Dirac}, the SLMC can greatly reduce the autocorrelation time in the conventional DQMC simulation and make the larger systems and lower temperature accessible. 

The other method is the elective momentum ultra-size quantum Monte Carlo (EQMC)~\cite{ZHLiu2018}. EQMC is inspired by the awareness that critical fluctuations mainly couples to fermions near the hot spots. Thus, instead of including all the fermion degrees of freedom, we ignore fermions far away from the hot spots and focus only on momentum points near the hot spots in the simulation. This approximation will produce different results for non-universal quantities compared with the original model, such as $h_c$ or critical temperature. However, for universal quantities, such as scaling exponents, which are independent of microscopic details and the high energy cutoff, EQMC has been shown to generate consistent values with those obtained from standard DQMC~\cite{ZHLiu2018}. 

In EQMC, because to a local coupling (in real space) becomes non-local in the momentum basis,
one can no longer use the local update as in standard DQMC, as that would cost $\beta N\cdot O(\beta N_f^3)$ computational complexity. Fortunately, there are cumulative update scheme in the SLMC developed recently~\cite{liu2016fermion,Xu2016self}. Such cumulative update is global move of the Ising spins and gives rise to the complexity $O(\beta N_f^3)$ for computing the fermion determinant. Since $N_f$ can be much smaller than $N$, speedup of the order $(\frac{N}{N_f})^3 \sim 10^3$ of EQMC over DQMC, with $\frac{N}{N_f} \sim 10$, can be easily achieved.

In the square lattice model, as shown in Fig.~\ref{fig:fig1} (b), the AFM wave vectors $\mathbf{Q}_i$ connect 4 pairs of hot spots ($N_{h.s.}=8$ in one layer and $N_{h.s.}=16$ in two).
In the IR limit, only fluctuations connecting each pair of hot spots are important to the universal scaling behavior in the vicinity of QCP~\cite{Metzner2003,Abanov2003,Abanov2004,Loehneysen2007,Metlitski2010,Metlitski2010a,Metlitski2010b}.
Hence, to study this universal behavior, we draw one patch around each $\mathbf{K}_{l}$ and keep fermion modes therein, and neglect other parts of the BZ.
In this way, instead of the original $N=L \times L$ momentum points, EQMC only keeps $N_f=L_f\times L_f$ momentum points for fermions inside each patch. Here, $L$ and $L_f$ denotes the linear size of the original lattice and the size of the patch, respectively.

\begin{figure}[htp]
	\centering
	\includegraphics[width=\columnwidth]{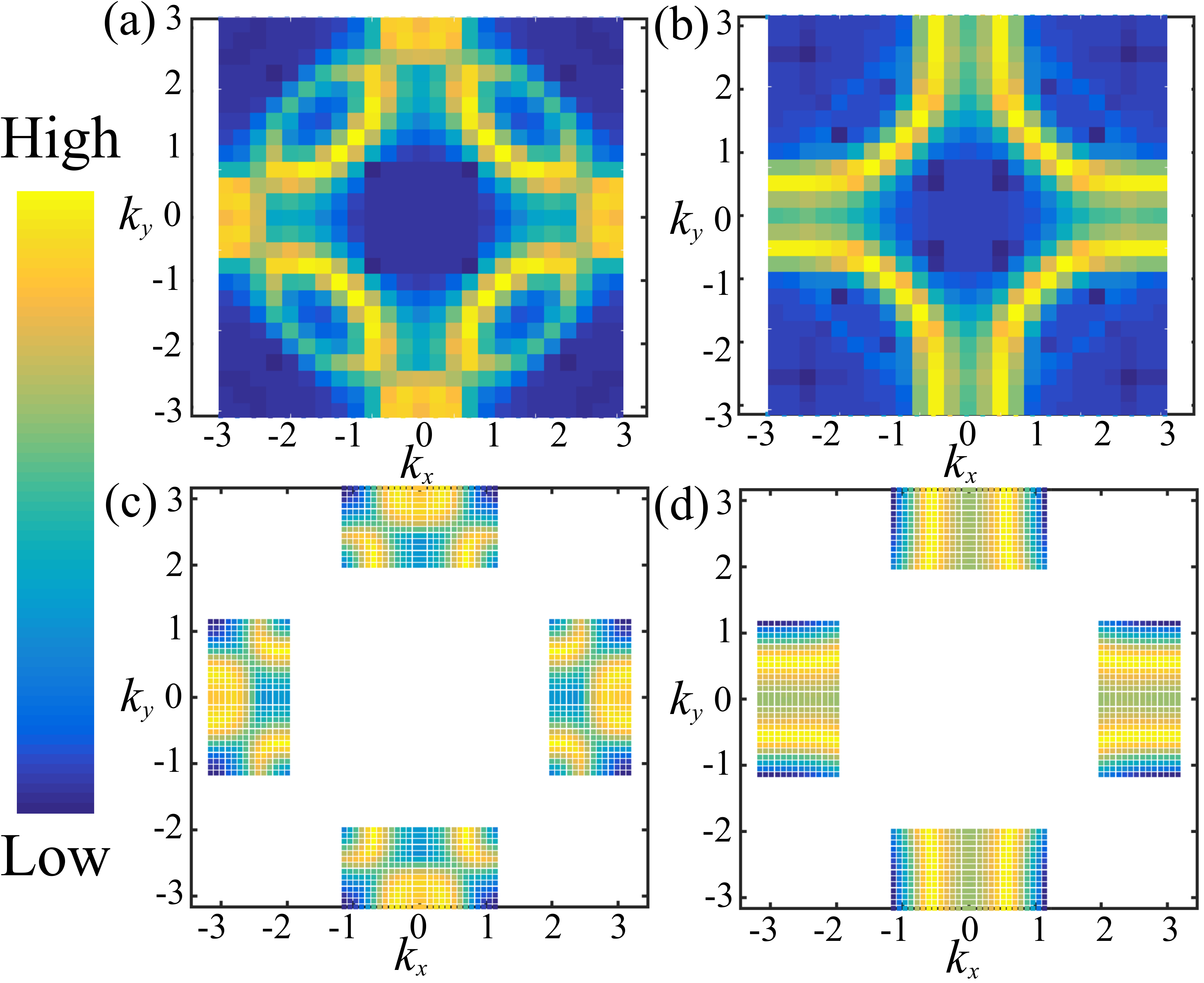}
	\caption{Ferm surface obtained from DQMC (panels (a) and (b)) and EQMC (panels (c) and (d)). Here we show the Fermi surface by plotting the fermion spectrum function at zero energy $A(\mathbf{k},\omega=0)$ utilizing the standard approximation $G(\mathbf{k},\beta/2) \sim A(\mathbf{k},\omega=0)$. (a) and (c), FS in the AFM ordered phase ($h<h_c$), where Fermi pockets are formed from zone folding. DQMC and EQMC results are consistent with each other, while EQMC (with system size $L=60$) gives much higher resolution in comparison with DQMC ($L=28$). (b) and (d), similar comparison at the QCP ($h=h_c$).}
	\label{fig:fig3}
\end{figure}

DQMC and EQMC are complementary to each other, the former provides unbiased results with relatively small systems and the latter, as an approximation, provides results closer to the QCP with finite size effects better suppressed. One other benefit of EQMC is that it provides much higher momentum resolution close to the hot spots. Fig.~\ref{fig:fig3} depicts the FS of the model in Eq.~\eqref{eq:HfM} obtained from $G(\mathbf{k},\beta/2) \sim A(\mathbf{k},\omega=0)$ via DQMC (panels (a) and (b)) and EQMC (panels (c) and (d)). The left panels are for $h<h_c$, i.e inside AFM metallic phase, whereas the right panels are for $h\sim h_c$, i.e. at the AFM-QCP. The DQMC data are obtained from $L=28$, $\beta=14$ simulations, it is clear that the momentum resolution is still too low to provided detailed FS structures near the hot spots. With EQMC, the system sizes are $L=60$ and $\beta=14$ in Fig.~\ref{fig:fig3} (c) and (d), and the momentum resolution is dramatically improved. For example, in Fig.~\ref{fig:fig3} (c), inside the AFM metallic phase, the gap at hot spots are clearly visualized. And in Fig.~\ref{fig:fig3} (d), at the AFM-QCP the FS recovers the shape of the non-interacting one, and non-Fermi-liquid behavior emerges at the hot spots as shown in the next section. To capture this important physics, EQMC and its higher momentum resolution play a vital role.

\section{Results}
\label{sec:results}

\subsection{Non-Fermi liquid}
\begin{figure}[t!]
	\centering
	\includegraphics[width=\columnwidth]{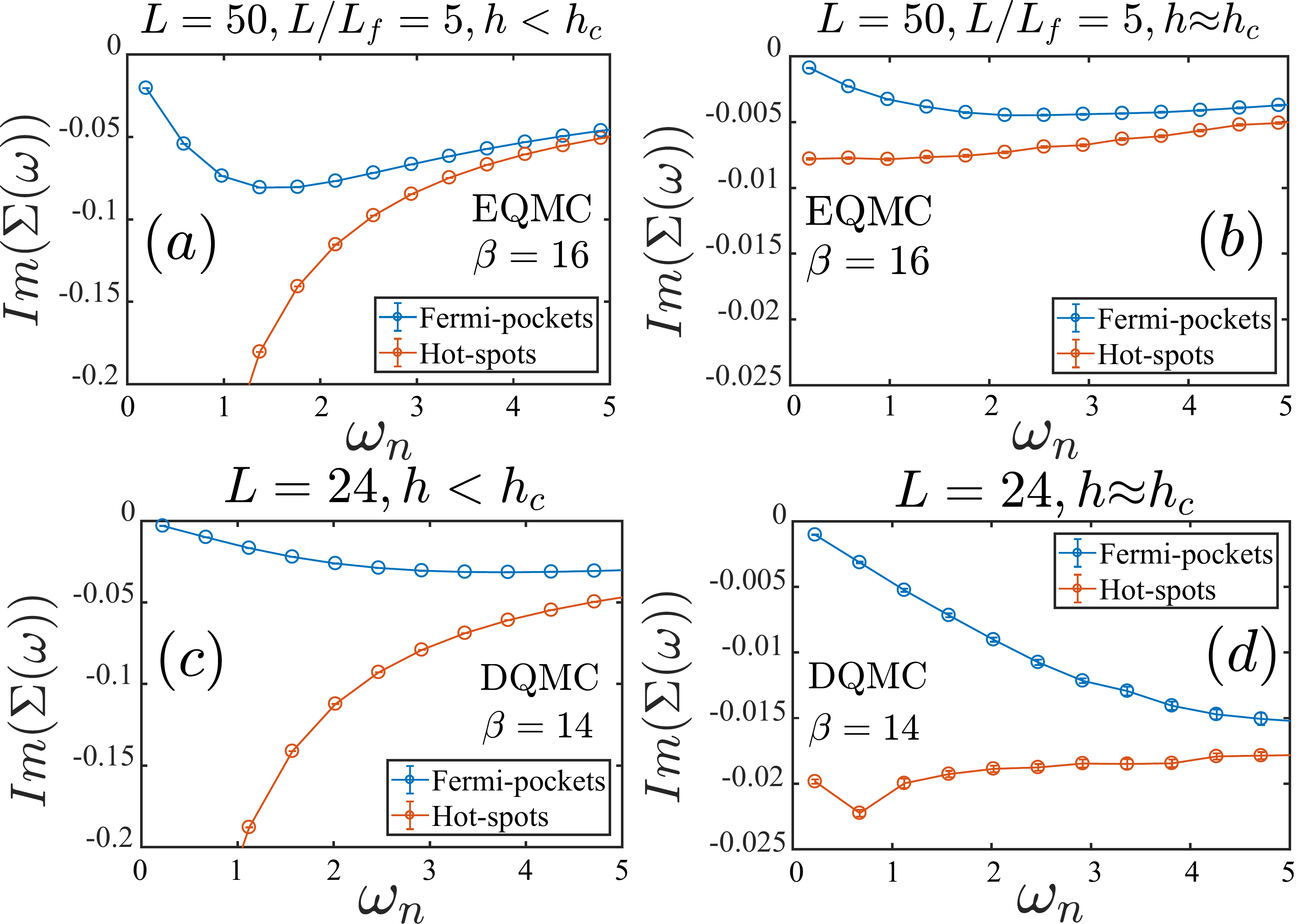}
	\caption{(a) Self-energy obtained from EQMC inside the AFM-metal phase ($h<h_c$). On the Fermi pockets, the system is in a Fermi liquid state as shown by the linearly vanishing of $\text{Im}(\Sigma(\mathbf{k},\omega))$ at small $\omega$. At the hot spot, due to the Fermi surface reconstruction, an energy gap opens up, resulting in a divergent $\text{Im}(\Sigma(\mathbf{k},\omega))$. (b) Self-energy obtained from EQMC at the AFM-QCP ($h\sim h_c$). On the Fermi pockets, the system remains a Fermi liquid as shown by the linearly vanishing $\text{Im}(\Sigma(\mathbf{k},\omega))$ at small $\omega$. At the hot spot, $\text{Im}(\Sigma(\mathbf{k},\omega))$ has a small but finite value as $\omega$ goes to 0, which is the signature of a non-Fermi-liquid. (c) and (d) show the same quantities produced in DQMC simulations with smaller system sizes, where qualitative behaviors remain the same.}
	\label{fig:fig4}
\end{figure}

As we emphasized above, the  dramatically improved momentum resolution in EQMC enables us to study the fermionic modes on the FS more precisely.
We studied the fermion self-energies in the AFM-metal phase and at the AFM-QCP. The results are shown in Fig.~\ref{fig:fig4}. 

In the AFM-metal phase, although the bands are folded accroding to Fig.~\ref{fig:fig1} (b), the system remains a Fermi liquid, with a band gap opening up at hot spots. Such expectations are revealed in Fig.~\ref{fig:fig4} (a) and (c). The Matsubara-frequency dependence of the $\text{Im}(\Sigma(\mathbf{k},\omega))$ either goes to zero linearly (on the pockets) or diverges (at the hot spot). Near the AFM-QCP, however, the situation is very different. Fermions at the hot spots shows non-Fermi liquid behavior, namely, as shown in Fig.~\ref{fig:fig4} (b) and (d), $\text{Im}(\Sigma(\mathbf{k},\omega))$ goes to a small constant at low $\omega$, and no sign of either vanishing or diverging are observed. The fermions away from the hot spots remains in Fermi-liquid like. Once again, DQMC and EQMC simulations give consistent results with the same qualitative behavior. 

It is worthwhile to point out, at the QCP, for fermions at the hot spot, a finite imaginary part in fermion self-energy is observed, which doesn't seem to decay to zero as we reduce the frequency. This behavior (a constant term in the imaginary part of the self-energy)  is not yet theoretically understood. However, it is consistent with similar QMC studies, where such a finite or constant term always seem to emerge near itinerant QCPs~\cite{Xu2017,ZHLiu2017,ZHLiu2018}.

\subsection{Universality class and critical exponents}
In our previous work on triangle lattice AFM-QCP~\cite{ZHLiu2017} with $2\mathbf{Q} \ne \Gamma$ (in fact in that case one has $3\mathbf{Q} = \Gamma$), the bosonic susceptibilities $\chi(T, h_c, \bm q, \omega)$ [as defined in Eq.~\eqref{eq:susceptibility} and Eq.~\eqref{eq:susceptibility_omega_k}] close to the QCP, revealed with $30\times 30\times 600$ ($L\times L\times L_{\tau}$) from DQMC, fits to the form of
\begin{align}
	& \chi(T,h_c,\mathbf{q},\omega_n) \nonumber \\
	& =\frac{1}{(c_{t}T+c'_{t}T^{2})+c_q |\mathbf{q}|^{2} + c_{\omega}\omega+c'_{\omega}\omega^2}.
	\label{eq:AFMTrisusceptibility}
\end{align}
In particular, at low $\omega$, $\chi^{-1}(0, h_c, 0, \omega)$ exhibits a crossover behavior from $\omega^2$ to $\omega$ and the susceptibility scales with $\mathbf{q}$ as $\chi^{-1}(0, h_c, \mathbf{q}, 0)\propto|\mathbf{q}|^2$, i.e., no anomalous dimension is observed. The system acquires dynamic critical exponent $z=2$, consistent with the Hertz-Millis mean-field expectation of the AFM-QCP at its upper critical dimension $d+z=4$.

\begin{figure}[t!]
	\centering
	\includegraphics[width=\columnwidth]{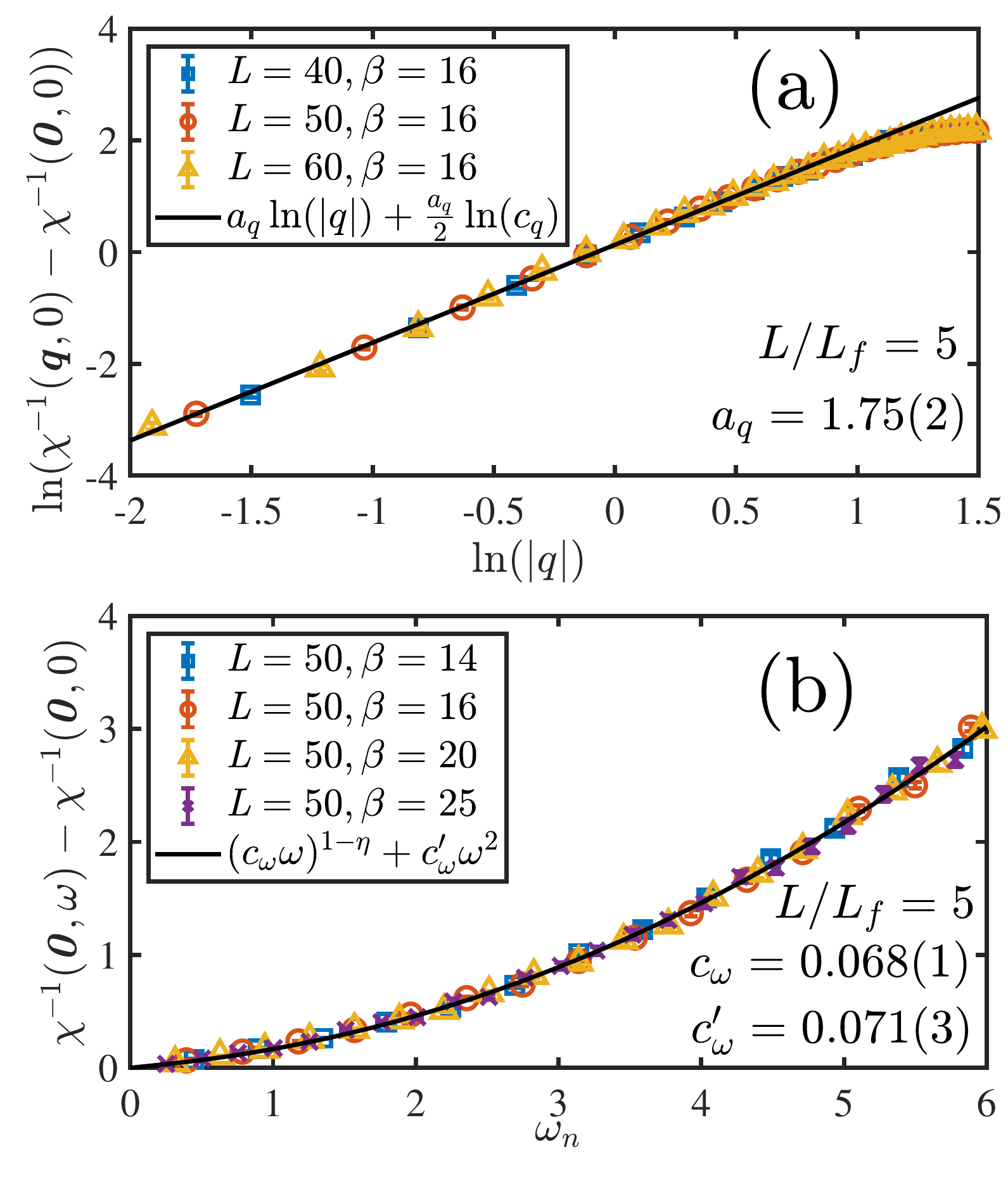}
	\caption{(a) $|\mathbf{q}|$ dependence of the bosonic susceptibilities $\chi(T=0,h=h_c,\mathbf{q},\omega=0)$ at the AFM-QCP. The system sizes are $L=40,50$ and $60$. The fitting line according to the form in Eq.~\eqref{eq:AFMQCPSQUARE} reveals that there is anomalous dimension in $\chi^{-1}(\mathbf{q})\sim |\mathbf{q}|^{2(1-\eta)}$ with $\eta=0.125$. (b) $\omega$ dependence of the bosonic susceptibilies $\chi(T=0,h=h_c,\mathbf{q}=0,\omega)$ at the AFM-QCP. The system size is $L=50$ and the temperature is as low as $\beta=25$ ($L_{\tau}=500$). The fitting line according to the form in Eq.~\eqref{eq:AFMQCPSQUARE} reveals that there is anomalous dimension in $\chi^{-1}(\omega) \sim \omega^{(1-\eta)}$ at small $\omega$ and crossover to $\chi^{-1}(\omega) \sim \omega^2$ at high $\omega$.}
	\label{fig:fig5}
\end{figure}

For the square lattice model in this paper, we expect the dynamic spin susceptibility has the following asymptotic form in the quantum critical region($h=h_c$)
\begin{align}
	& \chi(T,h_c,\mathbf{q},\omega_n) \nonumber \\
	& =\frac{1}{c_{t}T^{a_t}+(c_q |\mathbf{q}|^{2} + c_{\omega}\omega)^{1-\eta}+c'_{\omega}\omega^{2}}.
	\label{eq:AFMQCPSQUARE}
\end{align}
This functional form is similar to the Hertz-Millis theory, but we allow an anomalous dimension ($\eta$) as a free fitting parameter. We used this functional form to guide our data analysis.

We first look at the $\mathbf{q}$ dependence of $\chi^{-1}$, as shown in Fig.~\ref{fig:fig5} (a), the momentum $|\mathbf{q}|$ is measured with respect to the hot spot $\mathbf{K}$. 
Here we plot the susceptibility data by substracting the finite temperature background as, $\chi^{-1}(T,h_c,|\mathbf{q}|,0)-\chi^{-1}(T,h_c,0,0)=c_q |\mathbf{q}|^{a_q}$, where $a_q=2(1-\eta)$, and fit the curve to obtain the coefficient $c_q$ and the anomalous dimension $\eta$, as shown by the solid line in Fig.~\ref{fig:fig5} (a). Utilizing EQMC, with the system size as large as $L=60$, the powerlaw behavior $\chi^{-1}(|\mathbf{q}|)\propto |\mathbf{q}|^{a_{q}}$ clearly manifests, with $c_q=1.04(1)$ and $a_q=2(1-\eta)=1.75(2)$ with $\eta=0.125$. 
In DQMC simulation, we observed the same exponents $\eta=0.11(2)$, with a little bit lower accuracy due to smaller system sizes, the DQMC results are shown in the Sec. II of Appendix. 

As for the frequency dependence in $\chi$, as shown in Fig.~\ref{fig:fig5} (b), we analyze the $\chi^{-1}(T,h_c,0,\omega)-\chi^{-1}(T,h_c,0,0)$, to subtract the finite temperature background and test the predicted anomalous dimension $\eta=0.125$ in Eq.~\eqref{eq:AFMQCPSQUARE} and the data points fit very well the expected functional form 
\begin{equation}
	\chi^{-1}(\omega)=(c_{\omega}\omega)^{1-\eta}+ c'_\omega \omega^{2},
\end{equation}
where we obtained the values of the coefficients $c_\omega=0.068(1)$ and $c'_\omega=0.071(3)$. It is worthwhile to note that the crossover behavior in $\chi^{-1}(\omega)$ is very interesting, since at low-frequency, the anomalous dimension in $\omega^{0.875}$ dominates bosonic susceptibility and this means that the coupling of the fermions with the critical bosons have changed the universality behavior from the bare $(2+1)$D Ising universality with $\eta=0.036$ to a new one of AFM-QCP with $\eta=0.125$. However, at high-frequency, where the coupling between fermions and bosons becomes irrelevant, the bare boson university comes back and the $\omega^{2}$-term dominates over the susceptibility, consistent with our observation of the bare boson susceptiblity. 
We also note that because the frequency dependence here is polluted by the IR irrelevant $\omega^2$ contributions, whose contribution is about 10$\%$ at $\omega \sim 0.25$, this frequency exponent has a lower accuracy, in comparison with the momentum dependence shown in Fig.~\ref{fig:fig5} (a). Although the data are consistent with dynamical exponent $z=2$, small corrections in the form of anomalous dimension in the dynamical exponent as predicted in Ref.~\cite{Metlitski2010b} cannot be excluded.

\begin{figure}[htp]
	\centering
	\includegraphics[width=\columnwidth]{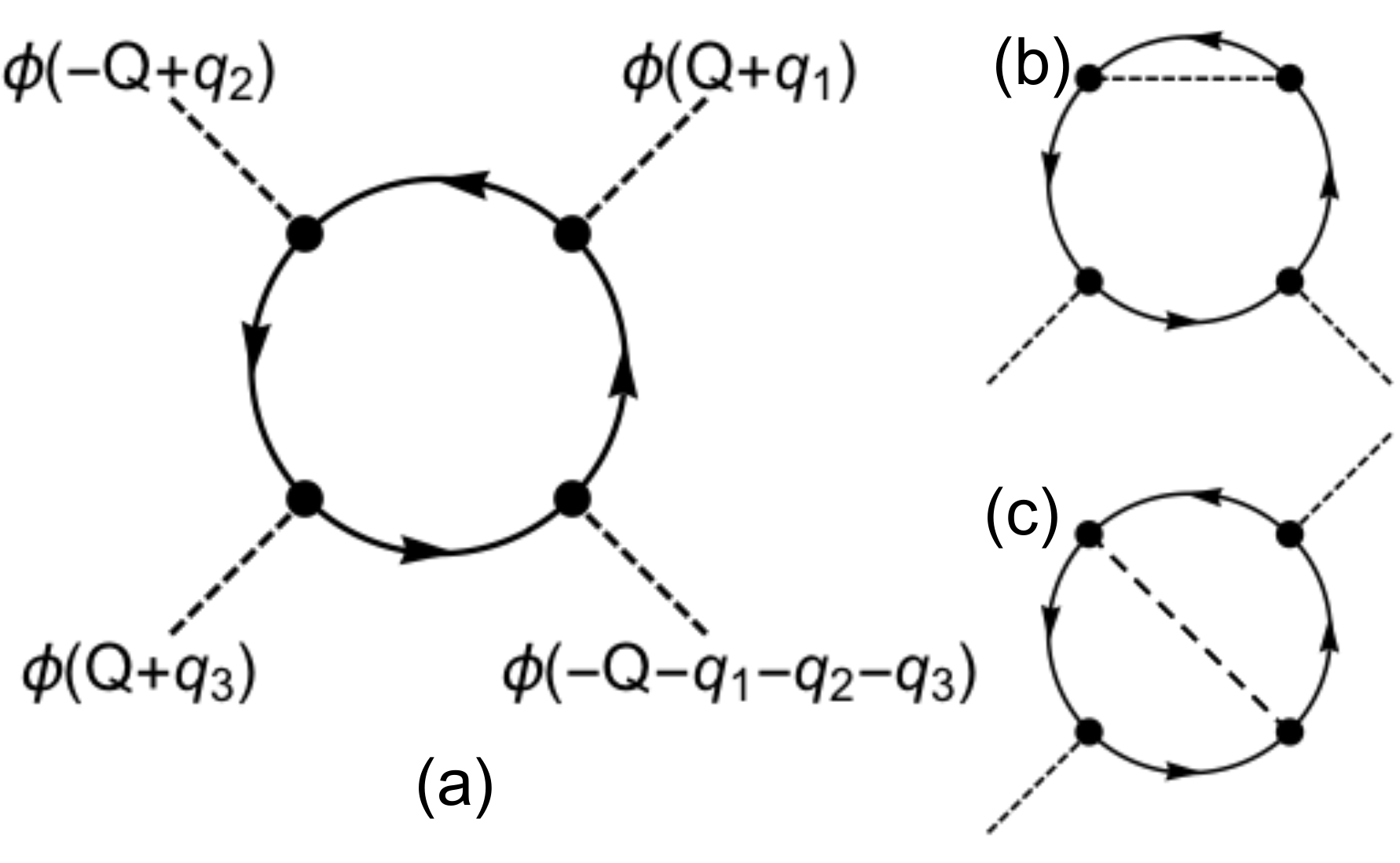}
	\caption{(a) Feynman diagram representing a four-boson interaction vertex. Dashed lines, $\phi(\mathbf{k})$, represent spin fluctuations at momentum $\mathbf{k}$ and we set $q<<Q$. Because low-energy physics is dominated by fermionic excitations near the FS, two of the four boson legs must have momenta near $\mathbf{Q}$, while the other two are near $-\mathbf{Q}$ to keep the fermions near the FS as shown in the figure. For $2\mathbf{Q}=\Gamma$, $+\mathbf{Q}$ and $-\mathbf{Q}$ becomes identical, and thus there exist two ways to contract the external legs as shown in (b) and (c). For $2\mathbf{Q} \ne \Gamma$, however, only the contraction shown in (b) is allowed, while the momentum conservation law is violated in (c).}
	\label{fig:fig6}
\end{figure}

In the absence of fermions, $\chi^{-1}(0,h_c,|\mathbf{q}|,0)\propto |\mathbf{q}|^{1.96}$ [$(2+1)$D Ising]. According to the Hertz-Millis theory, this exponent should increase from 1.96 to $2$ in the presence of fermions $\chi^{-1}(0,h_c,\mathbf{q},0)\propto |\mathbf{q}|^{2}$. Such an increase is indeed observed in triangular lattice model ($2 \mathbf{Q}\ne \Gamma$)~\cite{ZHLiu2017}. However, it is remarkable for the square lattice model ($2 \mathbf{Q}=\Gamma$), exactly the opposite was observed. Instead of increasing, this power actually decreases from $1.96$ to $1.75$,  $\chi^{-1}(0,h_c,\mathbf{q},0)\propto |\mathbf{q}|^{1.75}$. Such a significant contrast is beyond numerical error, and it indicates that QCPs with $2 \mathbf{Q}\ne \Gamma$ and $2 \mathbf{Q}=\Gamma$ belong to totally different universality classes, which is one of the key observation in our study.

This difference can be understand in the following way. Between $2 \mathbf{Q}=\Gamma$ and $2 \mathbf{Q}\ne \Gamma$, the constraints that the momentum conservation law enforces are different. As shown in Ref.~\cite{Abanov2003}, the QCPs with $2 \mathbf{Q}=\Gamma$ deviates from the Hertz-Millis theory already at the the level of four-boson vertex correction, as shown in Fig.~\ref{fig:fig6} (a). For $2 \mathbf{Q}=\Gamma$, this four-boson vertex gives two topologically different bosonic self-energy diagrams, Fig.~\ref{fig:fig6} (b) and (c), and in particular the diagram shown in Fig.~\ref{fig:fig6}(c) results in logarithmic corrections and is responsible for the breakdown of the Hertz-Millis scaling. However, for $2 \mathbf{Q}\ne\Gamma$ (e.g. in the triangular lattice model, we have $3\mathbf{Q}=\Gamma$ instead), this crucial diagram is prohibited by the momentum conservation law, and thus, at least within the same level of approximation, deviations from the Hertz-Millis picture is not expected. Further investigations, both analytical and numerical, are needed to better understand the role of this subtle difference, as well as the RG flows in other cases like $3\mathbf{Q}=\Gamma$, etc. 

\subsection{Comparison with RG analysis}
On the theory side, perturbative renormalization group calculation has been performed for Heisenberg AFM QCPs with SU(2) symmetry~\cite{Abanov2003,Metlitski2010b}, while the same study for Ising spins has not yet been carefully analyzed to our best knowledge. However, because some of the key features in the RG analysis are insensitive of the spin symmetry~\cite{Abanov2003,Metlitski2010b}, many qualitative results will hold and thus here we compare our numerical results with existing theoretical predictions for Heisenberg AFM QCPs, but it must also be emphasized that agreement at the quantitative level is not expected here because of this difference in symmetry. 

In the perturbative renormalization group calculation~\cite{Abanov2003,Metlitski2010b}, the anomalous dimension depends on the angle between the hot spot Fermi velocity and the order wavevector $\vec{Q}$. As shown below, for our model, this angle is close to $45^\circ$. At this angle, the RG-prediction for the anomalous dimension is $\eta=1/N_{h.s.}$ where $N_{h.s.}$ is the number of hot spots~\cite{Abanov2003,Metlitski2010b}. In our model, $N_{h.s.}=16$, and thus the RG predicted value is $\eta=1/16$. However, as shown above, although we observed the same qualitative behavior, the value of $\eta$ that we observed is close to $2/N_{h.s.}$ instead. Whether this quantitative disagreement is due to the symmetry difference (Heisenberg vs Ising) or some other contributions is an interesting open question. 

In addition, the RG analysis also predicts that near the QCP the Fermi surface at hot spots shall rotate towards nesting~\cite{Abanov2003}. This rotation of Fermi surface will further increases the anomalous dimension and can even renormalize the value of the dynamic critical exponent $z$~\cite{Abanov2003,Metlitski2010b}. As will be shown below, our study indeed observed this Fermi surface rotation near the QCP. However, because this RG flow is very slow, our hot-spot Fermi surface only rotated by about half degree in our simulation before stoped by cut-off. For such a small rotation, the resuling increase of anomalous dimension and the change in dynamic critical exponent  is too weak to be observed.


\begin{figure}[t!]
	\centering
	\includegraphics[width=\columnwidth]{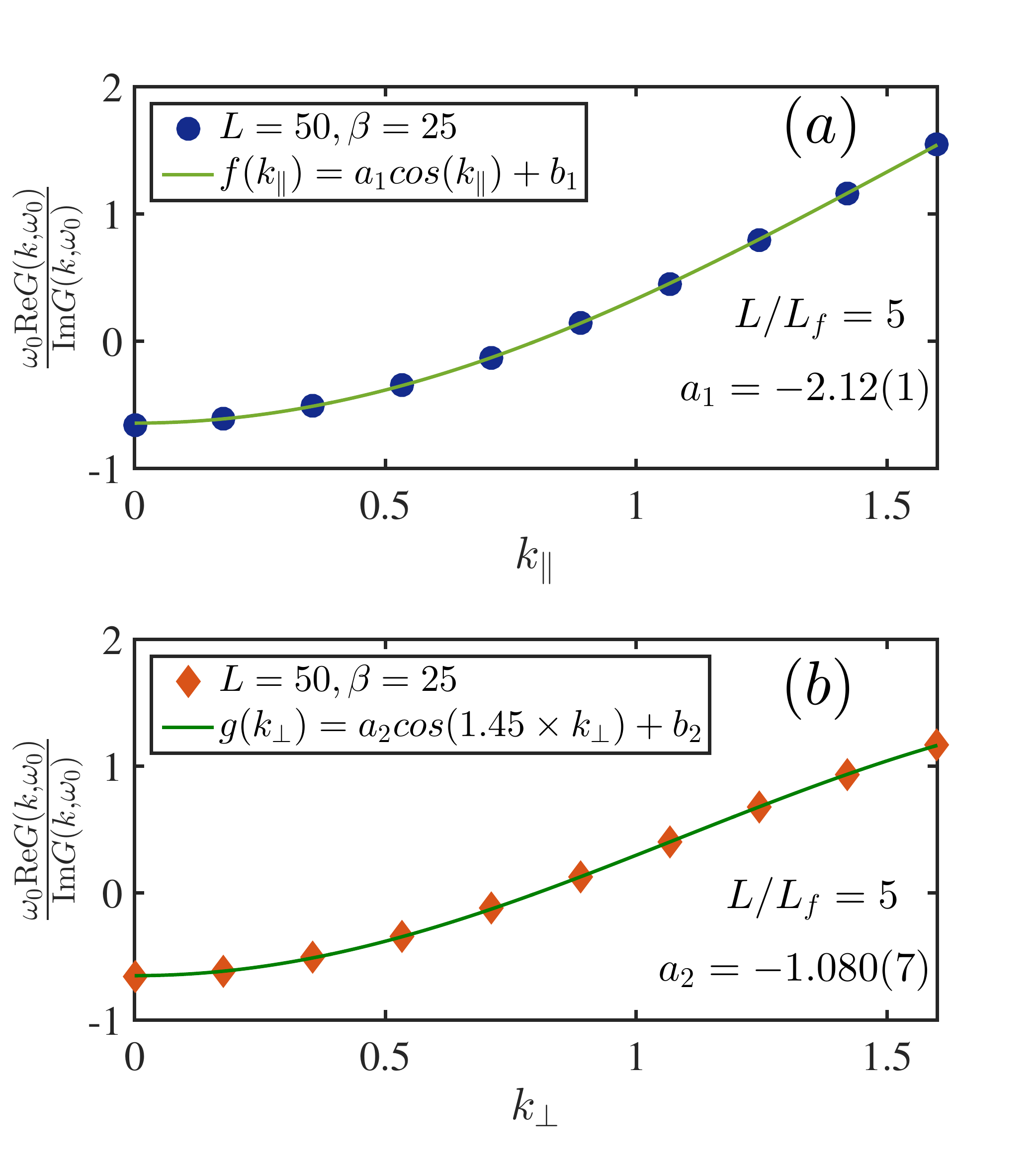}
	\caption{$\frac{\omega_0 \mathrm{Re}G(k,\omega_0)}{\mathrm{Im}G(k,\omega_0)}$ variation along (a) $Q=(\pi,\pi)$ direction $k_{\parallel}$ and (b) perpendicular to $Q=(\pi,\pi)$ direction $k_{\bot}$ at $K'_3$ hot spot mesh showed at Fig.~\ref{fig:fig1}. We use the simple trigonometric function $f(k_{\parallel})$ and $g(k_{\bot})$ to fit the data in (a) and (b). }
	\label{fig:fig7}
\end{figure} 

\begin{table}[ht]
	\caption{The Fermi velocity $v_F$ at hot spot in $K'_3$ hot spot mesh obtained from $\frac{\omega_0 \mathrm{Re}G(k,\omega_0)}{\mathrm{Im}G(k,\omega_0)}$ data showed in Fig.~\ref{fig:fig7} and the Fermi velocity in free fermion case.} 
	\centering 
	\def\arraystretch{1.5}
	\begin{tabular}{c c c}
		\hline\hline 
		Hot spots location & $k_x$ & $k_y$ \\[0.0ex]
		\hline
		& 2.5800 & 0.5615 \\[0.0ex]
		\hline\hline
		$v_F$ at hot spots & $v_{\parallel}$ & $v_{\bot}$ \\ [0.0ex] 
		\hline 
		Near QCP  &  1.523(8)  &  1.435(8)  \\ [0.0ex] 
		\hline                      
		Free fermion  &  1.506  &  1.468  \\ [0.0ex] 
		\hline 
	\end{tabular}
	\label{table1} 
\end{table}

We calculate the Fermi velocity $v_F$ as 
\begin{equation}
	\label{eq:Vf}
	\textbf{v}_F=\left.\frac{\partial}{\partial_{\textbf{k}}}\frac{\omega_0 \mathrm{Re}G(k,\omega_0)}{\mathrm{Im}G(k,\omega_0)}\right |_{\textbf{k}=\textbf{k}_F},
\end{equation} 
where $\omega_0=\pi/\beta$. To accurately compute the derivative, we first use simple functions to fit the discrete data points of $\frac{\omega_0 \mathrm{Re}G(k,\omega_0)}{\mathrm{Im}G(k,\omega_0)}$ vs $\mathbf{k}$ as showed in Fig.~\ref{fig:fig7}. And then, we compute the derivative for the fitting function to obtain the Fermi velocity, which is recorded in Table~\ref{table1}. $v_{\parallel}$ and $v_{\bot}$ are the two components of the Fermi velocity at a hot-spot parallel and perpendicular to $\mathbf{Q}$ respectively. In the table, we showed both the non-interacting Fermi velocity (bare values) as well as the Fermi velocity measured at the QCP (renormalized values).

According to the RG analysis~\cite{Abanov2003}, $v_{\parallel}$ and $v_{\bot}$  shall flow to infinite and zero respectively, but in the same time their product $v_{\parallel} \times v_{\bot}$ shall remain a constant. In a numerical simulation, this RG flow will be stoped by numerical cutoffs, e.g. finite size effects. Because this RG flow is marginal at tree level, the flow is expected to be very slow (logarithmic) and thus our observed renormalized value shall not differ dramatically from the bare ones. As can be seen from Table~\ref{table1}, this is indeed what we observed. The renormalized value of $v_{\parallel}$ ($v_{\bot}$) is slightly larger (smaller) than its bare value, and the product of $v_{\parallel}$ and $v_{\bot}$ remains largely a constant ($2.210$ for the bare values and $2.186$ for the renormalized ones), as the RG theory predicts. It is worthwhile to highlight here that although changes in $v_{\parallel}$ and $v_{\bot}$ are small, it is beyond numerical error as shown in Table~\ref{table1}, and theoretically, this small change in consistent with the slow RG flow predicted by theory.

\section*{Discussions}
\label{sec:conclusion}
If we compare the AFM-QCPs with $2 \mathbf{Q}\ne \Gamma$ and $2\mathbf{Q}=\Gamma$, QMC studies indicate that they belong to two different universality class, in contrast to the Hertz-Millis prediction, which doesn't rely on the value of $\mathbf{Q}$. This observation supports the $1/N_{h.s.}$ expansion and R.G. analysis discussed in Ref.~\cite{Abanov2003}.

In the study of criticality and anomalous critical scalings, the comparison between theory and numerical results plays a vital role. For QCPs in itinerant fermion systems,  although non-mean-field scaling beyond the Hertz-Millis theory has been predicted in theory and observed in QMC simulations, it has been a long standing challenge to reconcile numerical and theoretical results. Our study offers a solid example where an agreement between theory and numerical simulations starts to emerge, which is one first step towards a full understanding about itinerant QCPs~\cite{XiaoYanXu2019_review}. In particular, to pin point its exactly value of the frequency exponent and to probe the predicted anomalous dynamical critical exponents~\cite{Metlitski2010b,Schlief2017,SSLee2018}, lower temperature and frequency range needs to be explored. As pointed out in Refs.~\cite{Chubukov2018JC}, future works along this line are highly desirable, and are actively being pursued by us~\cite{XiaoYanXu2019_review}.

At the techinical level, a combination of DQMC and EQMC methodologies in this work shows a very promising direction in the numerical investigations of itinerant QCPs. Besides the consistency check in Ref.~\cite{ZHLiu2018} for triangular lattice AFM-QCP, the square lattice AFM-QCP investigated here provide the second example of the consistency in DQMC and EQMC in terms of revealing critical properties. Such consistency suggests a new pathway for future studies about quantum criticality in fermionic systems, in that, one can use DQMC on small systems to provide benchmark results, and the utilize EQMC to reveal IR physics at the thermodynamic limit.

\section*{Acknowledgement}
We acknowledge valuable discussions with Avraham Klein, Sung-Sik Lee, Yoni Schattner, Andrey Chubukov, Subir Sachdev and Steven Kivelson on various subjects of itinerant quantum critcality. ZHL, GPP and ZYM acknowledge fundings from the Ministry of Science and Technology of China through the National Key Research and Development Program (2016YFA0300502), the Strategic Priority Research Program of the Chinese Academy of Sciences (XDB28000000) and from the National Science Foundation of China under Grant Nos. 11421092, 11574359 and 11674370. XYX is thankful for the support of HKRGC through grant C6026-16W. K.S. acknowledges support from the National Science Foundation under Grant No. EFRI-1741618 and the Alfred P. Sloan Foundation. We thank the Center for Quantum Simulation Sciences in the Institute of Physics, Chinese Academy of Sciences and the Tianhe-1A and Tianhe-II platforms at the National Supercomputer Centers in Tianjin and Guangzhou for their technical support and generous allocation of CPU time.

\bibliography{AFMQCP.bib}

\begin{thebibliography}{66}%
\makeatletter
\providecommand \@ifxundefined [1]{%
 \@ifx{#1\undefined}
}%
\providecommand \@ifnum [1]{%
 \ifnum #1\expandafter \@firstoftwo
 \else \expandafter \@secondoftwo
 \fi
}%
\providecommand \@ifx [1]{%
 \ifx #1\expandafter \@firstoftwo
 \else \expandafter \@secondoftwo
 \fi
}%
\providecommand \natexlab [1]{#1}%
\providecommand \enquote  [1]{``#1''}%
\providecommand \bibnamefont  [1]{#1}%
\providecommand \bibfnamefont [1]{#1}%
\providecommand \citenamefont [1]{#1}%
\providecommand \href@noop [0]{\@secondoftwo}%
\providecommand \href [0]{\begingroup \@sanitize@url \@href}%
\providecommand \@href[1]{\@@startlink{#1}\@@href}%
\providecommand \@@href[1]{\endgroup#1\@@endlink}%
\providecommand \@sanitize@url [0]{\catcode `\\12\catcode `\$12\catcode
  `\&12\catcode `\#12\catcode `\^12\catcode `\_12\catcode `\%12\relax}%
\providecommand \@@startlink[1]{}%
\providecommand \@@endlink[0]{}%
\providecommand \url  [0]{\begingroup\@sanitize@url \@url }%
\providecommand \@url [1]{\endgroup\@href {#1}{\urlprefix }}%
\providecommand \urlprefix  [0]{URL }%
\providecommand \Eprint [0]{\href }%
\providecommand \doibase [0]{http://dx.doi.org/}%
\providecommand \selectlanguage [0]{\@gobble}%
\providecommand \bibinfo  [0]{\@secondoftwo}%
\providecommand \bibfield  [0]{\@secondoftwo}%
\providecommand \translation [1]{[#1]}%
\providecommand \BibitemOpen [0]{}%
\providecommand \bibitemStop [0]{}%
\providecommand \bibitemNoStop [0]{.\EOS\space}%
\providecommand \EOS [0]{\spacefactor3000\relax}%
\providecommand \BibitemShut  [1]{\csname bibitem#1\endcsname}%
\let\auto@bib@innerbib\@empty
\bibitem [{\citenamefont {Hertz}(1976)}]{Hertz1976}%
  \BibitemOpen
  \bibfield  {author} {\bibinfo {author} {\bibfnamefont {J.~A.}\ \bibnamefont
  {Hertz}},\ }\href {\doibase 10.1103/PhysRevB.14.1165} {\bibfield  {journal}
  {\bibinfo  {journal} {Phys. Rev. B}\ }\textbf {\bibinfo {volume} {14}},\
  \bibinfo {pages} {1165} (\bibinfo {year} {1976})}\BibitemShut {NoStop}%
\bibitem [{\citenamefont {Millis}(1993)}]{Millis1993}%
  \BibitemOpen
  \bibfield  {author} {\bibinfo {author} {\bibfnamefont {A.~J.}\ \bibnamefont
  {Millis}},\ }\href {\doibase 10.1103/PhysRevB.48.7183} {\bibfield  {journal}
  {\bibinfo  {journal} {Phys. Rev. B}\ }\textbf {\bibinfo {volume} {48}},\
  \bibinfo {pages} {7183} (\bibinfo {year} {1993})}\BibitemShut {NoStop}%
\bibitem [{\citenamefont {Moriya}(1985)}]{Moriya1985}%
  \BibitemOpen
  \bibfield  {author} {\bibinfo {author} {\bibfnamefont {T.}~\bibnamefont
  {Moriya}},\ }\href {\doibase 10.1007/978-3-642-82499-9} {\emph {\bibinfo
  {title} {Spin Fluctuations in Itinerant Electron Magnetism}}}\ (\bibinfo
  {publisher} {Springer-Verlag Berlin Heidelberg},\ \bibinfo {year}
  {1985})\BibitemShut {NoStop}%
\bibitem [{\citenamefont {Stewart}(2001)}]{Stewart2001}%
  \BibitemOpen
  \bibfield  {author} {\bibinfo {author} {\bibfnamefont {G.~R.}\ \bibnamefont
  {Stewart}},\ }\href {\doibase 10.1103/RevModPhys.73.797} {\bibfield
  {journal} {\bibinfo  {journal} {Rev. Mod. Phys.}\ }\textbf {\bibinfo {volume}
  {73}},\ \bibinfo {pages} {797} (\bibinfo {year} {2001})}\BibitemShut
  {NoStop}%
\bibitem [{\citenamefont {Chubukov}\ \emph {et~al.}(2004)\citenamefont
  {Chubukov}, \citenamefont {P\'epin},\ and\ \citenamefont
  {Rech}}]{Chubukov2004}%
  \BibitemOpen
  \bibfield  {author} {\bibinfo {author} {\bibfnamefont {A.~V.}\ \bibnamefont
  {Chubukov}}, \bibinfo {author} {\bibfnamefont {C.}~\bibnamefont {P\'epin}}, \
  and\ \bibinfo {author} {\bibfnamefont {J.}~\bibnamefont {Rech}},\ }\href
  {\doibase 10.1103/PhysRevLett.92.147003} {\bibfield  {journal} {\bibinfo
  {journal} {Phys. Rev. Lett.}\ }\textbf {\bibinfo {volume} {92}},\ \bibinfo
  {pages} {147003} (\bibinfo {year} {2004})}\BibitemShut {NoStop}%
\bibitem [{\citenamefont {Belitz}\ \emph {et~al.}(2005)\citenamefont {Belitz},
  \citenamefont {Kirkpatrick},\ and\ \citenamefont {Vojta}}]{Belitz2005}%
  \BibitemOpen
  \bibfield  {author} {\bibinfo {author} {\bibfnamefont {D.}~\bibnamefont
  {Belitz}}, \bibinfo {author} {\bibfnamefont {T.~R.}\ \bibnamefont
  {Kirkpatrick}}, \ and\ \bibinfo {author} {\bibfnamefont {T.}~\bibnamefont
  {Vojta}},\ }\href {\doibase 10.1103/RevModPhys.77.579} {\bibfield  {journal}
  {\bibinfo  {journal} {Rev. Mod. Phys.}\ }\textbf {\bibinfo {volume} {77}},\
  \bibinfo {pages} {579} (\bibinfo {year} {2005})}\BibitemShut {NoStop}%
\bibitem [{\citenamefont {L\"ohneysen}\ \emph {et~al.}(2007)\citenamefont
  {L\"ohneysen}, \citenamefont {Rosch}, \citenamefont {Vojta},\ and\
  \citenamefont {W\"olfle}}]{Loehneysen2007}%
  \BibitemOpen
  \bibfield  {author} {\bibinfo {author} {\bibfnamefont {H.~v.}\ \bibnamefont
  {L\"ohneysen}}, \bibinfo {author} {\bibfnamefont {A.}~\bibnamefont {Rosch}},
  \bibinfo {author} {\bibfnamefont {M.}~\bibnamefont {Vojta}}, \ and\ \bibinfo
  {author} {\bibfnamefont {P.}~\bibnamefont {W\"olfle}},\ }\href {\doibase
  10.1103/RevModPhys.79.1015} {\bibfield  {journal} {\bibinfo  {journal} {Rev.
  Mod. Phys.}\ }\textbf {\bibinfo {volume} {79}},\ \bibinfo {pages} {1015}
  (\bibinfo {year} {2007})}\BibitemShut {NoStop}%
\bibitem [{\citenamefont {Chubukov}\ and\ \citenamefont
  {Maslov}(2009)}]{Chubukov2009}%
  \BibitemOpen
  \bibfield  {author} {\bibinfo {author} {\bibfnamefont {A.~V.}\ \bibnamefont
  {Chubukov}}\ and\ \bibinfo {author} {\bibfnamefont {D.~L.}\ \bibnamefont
  {Maslov}},\ }\href {\doibase 10.1103/PhysRevLett.103.216401} {\bibfield
  {journal} {\bibinfo  {journal} {Phys. Rev. Lett.}\ }\textbf {\bibinfo
  {volume} {103}},\ \bibinfo {pages} {216401} (\bibinfo {year}
  {2009})}\BibitemShut {NoStop}%
\bibitem [{\citenamefont {Metzner}\ \emph {et~al.}(2003)\citenamefont
  {Metzner}, \citenamefont {Rohe},\ and\ \citenamefont
  {Andergassen}}]{Metzner2003}%
  \BibitemOpen
  \bibfield  {author} {\bibinfo {author} {\bibfnamefont {W.}~\bibnamefont
  {Metzner}}, \bibinfo {author} {\bibfnamefont {D.}~\bibnamefont {Rohe}}, \
  and\ \bibinfo {author} {\bibfnamefont {S.}~\bibnamefont {Andergassen}},\
  }\href {\doibase 10.1103/PhysRevLett.91.066402} {\bibfield  {journal}
  {\bibinfo  {journal} {Phys. Rev. Lett.}\ }\textbf {\bibinfo {volume} {91}},\
  \bibinfo {pages} {066402} (\bibinfo {year} {2003})}\BibitemShut {NoStop}%
\bibitem [{\citenamefont {Senthil}(2008)}]{Senthil2008}%
  \BibitemOpen
  \bibfield  {author} {\bibinfo {author} {\bibfnamefont {T.}~\bibnamefont
  {Senthil}},\ }\href {\doibase 10.1103/PhysRevB.78.035103} {\bibfield
  {journal} {\bibinfo  {journal} {Phys. Rev. B}\ }\textbf {\bibinfo {volume}
  {78}},\ \bibinfo {pages} {035103} (\bibinfo {year} {2008})}\BibitemShut
  {NoStop}%
\bibitem [{\citenamefont {Holder}\ and\ \citenamefont
  {Metzner}(2015)}]{Holder2015}%
  \BibitemOpen
  \bibfield  {author} {\bibinfo {author} {\bibfnamefont {T.}~\bibnamefont
  {Holder}}\ and\ \bibinfo {author} {\bibfnamefont {W.}~\bibnamefont
  {Metzner}},\ }\href {\doibase 10.1103/PhysRevB.92.041112} {\bibfield
  {journal} {\bibinfo  {journal} {Phys. Rev. B}\ }\textbf {\bibinfo {volume}
  {92}},\ \bibinfo {pages} {041112} (\bibinfo {year} {2015})}\BibitemShut
  {NoStop}%
\bibitem [{\citenamefont {Metlitski}\ \emph {et~al.}(2015)\citenamefont
  {Metlitski}, \citenamefont {Mross}, \citenamefont {Sachdev},\ and\
  \citenamefont {Senthil}}]{Metlitski2015}%
  \BibitemOpen
  \bibfield  {author} {\bibinfo {author} {\bibfnamefont {M.~A.}\ \bibnamefont
  {Metlitski}}, \bibinfo {author} {\bibfnamefont {D.~F.}\ \bibnamefont
  {Mross}}, \bibinfo {author} {\bibfnamefont {S.}~\bibnamefont {Sachdev}}, \
  and\ \bibinfo {author} {\bibfnamefont {T.}~\bibnamefont {Senthil}},\ }\href
  {\doibase 10.1103/PhysRevB.91.115111} {\bibfield  {journal} {\bibinfo
  {journal} {Phys. Rev. B}\ }\textbf {\bibinfo {volume} {91}},\ \bibinfo
  {pages} {115111} (\bibinfo {year} {2015})}\BibitemShut {NoStop}%
\bibitem [{\citenamefont {Xu}\ \emph {et~al.}(2017{\natexlab{a}})\citenamefont
  {Xu}, \citenamefont {Sun}, \citenamefont {Schattner}, \citenamefont {Berg},\
  and\ \citenamefont {Meng}}]{Xu2017}%
  \BibitemOpen
  \bibfield  {author} {\bibinfo {author} {\bibfnamefont {X.~Y.}\ \bibnamefont
  {Xu}}, \bibinfo {author} {\bibfnamefont {K.}~\bibnamefont {Sun}}, \bibinfo
  {author} {\bibfnamefont {Y.}~\bibnamefont {Schattner}}, \bibinfo {author}
  {\bibfnamefont {E.}~\bibnamefont {Berg}}, \ and\ \bibinfo {author}
  {\bibfnamefont {Z.~Y.}\ \bibnamefont {Meng}},\ }\href {\doibase
  10.1103/PhysRevX.7.031058} {\bibfield  {journal} {\bibinfo  {journal} {Phys.
  Rev. X}\ }\textbf {\bibinfo {volume} {7}},\ \bibinfo {pages} {031058}
  (\bibinfo {year} {2017}{\natexlab{a}})}\BibitemShut {NoStop}%
\bibitem [{\citenamefont {Custers}\ \emph {et~al.}(2003)\citenamefont
  {Custers}, \citenamefont {Gegenwart}, \citenamefont {Wilhelm}, \citenamefont
  {Neumaier}, \citenamefont {Tokiwa}, \citenamefont {Trovarelli}, \citenamefont
  {Geibel}, \citenamefont {Steglich}, \citenamefont {Pepin},\ and\
  \citenamefont {Coleman}}]{Custers2003}%
  \BibitemOpen
  \bibfield  {author} {\bibinfo {author} {\bibfnamefont {J.}~\bibnamefont
  {Custers}}, \bibinfo {author} {\bibfnamefont {P.}~\bibnamefont {Gegenwart}},
  \bibinfo {author} {\bibfnamefont {H.}~\bibnamefont {Wilhelm}}, \bibinfo
  {author} {\bibfnamefont {K.}~\bibnamefont {Neumaier}}, \bibinfo {author}
  {\bibfnamefont {Y.}~\bibnamefont {Tokiwa}}, \bibinfo {author} {\bibfnamefont
  {O.}~\bibnamefont {Trovarelli}}, \bibinfo {author} {\bibfnamefont
  {C.}~\bibnamefont {Geibel}}, \bibinfo {author} {\bibfnamefont
  {F.}~\bibnamefont {Steglich}}, \bibinfo {author} {\bibfnamefont
  {C.}~\bibnamefont {Pepin}}, \ and\ \bibinfo {author} {\bibfnamefont
  {P.}~\bibnamefont {Coleman}},\ }\href {http://dx.doi.org/10.1038/nature01774}
  {\bibfield  {journal} {\bibinfo  {journal} {Nature}\ }\textbf {\bibinfo
  {volume} {424}},\ \bibinfo {pages} {524 } (\bibinfo {year}
  {2003})}\BibitemShut {NoStop}%
\bibitem [{\citenamefont {Steppke}\ \emph {et~al.}(2013)\citenamefont
  {Steppke}, \citenamefont {K{\"u}chler}, \citenamefont {Lausberg},
  \citenamefont {Lengyel}, \citenamefont {Steinke}, \citenamefont {Borth},
  \citenamefont {L{\"u}hmann}, \citenamefont {Krellner}, \citenamefont
  {Nicklas}, \citenamefont {Geibel}, \citenamefont {Steglich},\ and\
  \citenamefont {Brando}}]{Steppke2013}%
  \BibitemOpen
  \bibfield  {author} {\bibinfo {author} {\bibfnamefont {A.}~\bibnamefont
  {Steppke}}, \bibinfo {author} {\bibfnamefont {R.}~\bibnamefont
  {K{\"u}chler}}, \bibinfo {author} {\bibfnamefont {S.}~\bibnamefont
  {Lausberg}}, \bibinfo {author} {\bibfnamefont {E.}~\bibnamefont {Lengyel}},
  \bibinfo {author} {\bibfnamefont {L.}~\bibnamefont {Steinke}}, \bibinfo
  {author} {\bibfnamefont {R.}~\bibnamefont {Borth}}, \bibinfo {author}
  {\bibfnamefont {T.}~\bibnamefont {L{\"u}hmann}}, \bibinfo {author}
  {\bibfnamefont {C.}~\bibnamefont {Krellner}}, \bibinfo {author}
  {\bibfnamefont {M.}~\bibnamefont {Nicklas}}, \bibinfo {author} {\bibfnamefont
  {C.}~\bibnamefont {Geibel}}, \bibinfo {author} {\bibfnamefont
  {F.}~\bibnamefont {Steglich}}, \ and\ \bibinfo {author} {\bibfnamefont
  {M.}~\bibnamefont {Brando}},\ }\href {\doibase 10.1126/science.1230583}
  {\bibfield  {journal} {\bibinfo  {journal} {Science}\ }\textbf {\bibinfo
  {volume} {339}},\ \bibinfo {pages} {933} (\bibinfo {year}
  {2013})}\BibitemShut {NoStop}%
\bibitem [{\citenamefont {Zhang}\ \emph {et~al.}(2016)\citenamefont {Zhang},
  \citenamefont {Park}, \citenamefont {Lu}, \citenamefont {Wei}, \citenamefont
  {Ma}, \citenamefont {Hao}, \citenamefont {Dai}, \citenamefont {Meng},
  \citenamefont {Yang}, \citenamefont {Luo},\ and\ \citenamefont
  {Li}}]{ZhangWenLiang2016}%
  \BibitemOpen
  \bibfield  {author} {\bibinfo {author} {\bibfnamefont {W.}~\bibnamefont
  {Zhang}}, \bibinfo {author} {\bibfnamefont {J.~T.}\ \bibnamefont {Park}},
  \bibinfo {author} {\bibfnamefont {X.}~\bibnamefont {Lu}}, \bibinfo {author}
  {\bibfnamefont {Y.}~\bibnamefont {Wei}}, \bibinfo {author} {\bibfnamefont
  {X.}~\bibnamefont {Ma}}, \bibinfo {author} {\bibfnamefont {L.}~\bibnamefont
  {Hao}}, \bibinfo {author} {\bibfnamefont {P.}~\bibnamefont {Dai}}, \bibinfo
  {author} {\bibfnamefont {Z.~Y.}\ \bibnamefont {Meng}}, \bibinfo {author}
  {\bibfnamefont {Y.-f.}\ \bibnamefont {Yang}}, \bibinfo {author}
  {\bibfnamefont {H.}~\bibnamefont {Luo}}, \ and\ \bibinfo {author}
  {\bibfnamefont {S.}~\bibnamefont {Li}},\ }\href {\doibase
  10.1103/PhysRevLett.117.227003} {\bibfield  {journal} {\bibinfo  {journal}
  {Phys. Rev. Lett.}\ }\textbf {\bibinfo {volume} {117}},\ \bibinfo {pages}
  {227003} (\bibinfo {year} {2016})}\BibitemShut {NoStop}%
\bibitem [{\citenamefont {Liu}\ \emph {et~al.}(2016)\citenamefont {Liu},
  \citenamefont {Gu}, \citenamefont {Zhang}, \citenamefont {Gong},
  \citenamefont {Zhang}, \citenamefont {Xie}, \citenamefont {Lu}, \citenamefont
  {Ma}, \citenamefont {Zhang}, \citenamefont {Zhang}, \citenamefont {Zhu},
  \citenamefont {Ren}, \citenamefont {Shan}, \citenamefont {Qiu}, \citenamefont
  {Dai}, \citenamefont {Yang}, \citenamefont {Luo},\ and\ \citenamefont
  {Li}}]{LiuZhaoYu2016}%
  \BibitemOpen
  \bibfield  {author} {\bibinfo {author} {\bibfnamefont {Z.}~\bibnamefont
  {Liu}}, \bibinfo {author} {\bibfnamefont {Y.}~\bibnamefont {Gu}}, \bibinfo
  {author} {\bibfnamefont {W.}~\bibnamefont {Zhang}}, \bibinfo {author}
  {\bibfnamefont {D.}~\bibnamefont {Gong}}, \bibinfo {author} {\bibfnamefont
  {W.}~\bibnamefont {Zhang}}, \bibinfo {author} {\bibfnamefont
  {T.}~\bibnamefont {Xie}}, \bibinfo {author} {\bibfnamefont {X.}~\bibnamefont
  {Lu}}, \bibinfo {author} {\bibfnamefont {X.}~\bibnamefont {Ma}}, \bibinfo
  {author} {\bibfnamefont {X.}~\bibnamefont {Zhang}}, \bibinfo {author}
  {\bibfnamefont {R.}~\bibnamefont {Zhang}}, \bibinfo {author} {\bibfnamefont
  {J.}~\bibnamefont {Zhu}}, \bibinfo {author} {\bibfnamefont {C.}~\bibnamefont
  {Ren}}, \bibinfo {author} {\bibfnamefont {L.}~\bibnamefont {Shan}}, \bibinfo
  {author} {\bibfnamefont {X.}~\bibnamefont {Qiu}}, \bibinfo {author}
  {\bibfnamefont {P.}~\bibnamefont {Dai}}, \bibinfo {author} {\bibfnamefont
  {Y.-f.}\ \bibnamefont {Yang}}, \bibinfo {author} {\bibfnamefont
  {H.}~\bibnamefont {Luo}}, \ and\ \bibinfo {author} {\bibfnamefont
  {S.}~\bibnamefont {Li}},\ }\href {\doibase 10.1103/PhysRevLett.117.157002}
  {\bibfield  {journal} {\bibinfo  {journal} {Phys. Rev. Lett.}\ }\textbf
  {\bibinfo {volume} {117}},\ \bibinfo {pages} {157002} (\bibinfo {year}
  {2016})}\BibitemShut {NoStop}%
\bibitem [{\citenamefont {Gu}\ \emph {et~al.}(2017)\citenamefont {Gu},
  \citenamefont {Liu}, \citenamefont {Xie}, \citenamefont {Zhang},
  \citenamefont {Gong}, \citenamefont {Hu}, \citenamefont {Ma}, \citenamefont
  {Li}, \citenamefont {Zhao}, \citenamefont {Lin}, \citenamefont {Xu},
  \citenamefont {Tan}, \citenamefont {Chen}, \citenamefont {Meng},
  \citenamefont {Yang}, \citenamefont {Luo},\ and\ \citenamefont
  {Li}}]{Gu2017}%
  \BibitemOpen
  \bibfield  {author} {\bibinfo {author} {\bibfnamefont {Y.}~\bibnamefont
  {Gu}}, \bibinfo {author} {\bibfnamefont {Z.}~\bibnamefont {Liu}}, \bibinfo
  {author} {\bibfnamefont {T.}~\bibnamefont {Xie}}, \bibinfo {author}
  {\bibfnamefont {W.}~\bibnamefont {Zhang}}, \bibinfo {author} {\bibfnamefont
  {D.}~\bibnamefont {Gong}}, \bibinfo {author} {\bibfnamefont {D.}~\bibnamefont
  {Hu}}, \bibinfo {author} {\bibfnamefont {X.}~\bibnamefont {Ma}}, \bibinfo
  {author} {\bibfnamefont {C.}~\bibnamefont {Li}}, \bibinfo {author}
  {\bibfnamefont {L.}~\bibnamefont {Zhao}}, \bibinfo {author} {\bibfnamefont
  {L.}~\bibnamefont {Lin}}, \bibinfo {author} {\bibfnamefont {Z.}~\bibnamefont
  {Xu}}, \bibinfo {author} {\bibfnamefont {G.}~\bibnamefont {Tan}}, \bibinfo
  {author} {\bibfnamefont {G.}~\bibnamefont {Chen}}, \bibinfo {author}
  {\bibfnamefont {Z.~Y.}\ \bibnamefont {Meng}}, \bibinfo {author}
  {\bibfnamefont {Y.-f.}\ \bibnamefont {Yang}}, \bibinfo {author}
  {\bibfnamefont {H.}~\bibnamefont {Luo}}, \ and\ \bibinfo {author}
  {\bibfnamefont {S.}~\bibnamefont {Li}},\ }\href {\doibase
  10.1103/PhysRevLett.119.157001} {\bibfield  {journal} {\bibinfo  {journal}
  {Phys. Rev. Lett.}\ }\textbf {\bibinfo {volume} {119}},\ \bibinfo {pages}
  {157001} (\bibinfo {year} {2017})}\BibitemShut {NoStop}%
\bibitem [{\citenamefont {Wu}\ \emph {et~al.}(2014)\citenamefont {Wu},
  \citenamefont {Cheng}, \citenamefont {Matsubayashi}, \citenamefont {Kong},
  \citenamefont {Lin}, \citenamefont {Jin}, \citenamefont {Wang}, \citenamefont
  {Uwatoko},\ and\ \citenamefont {Luo}}]{Wu2014}%
  \BibitemOpen
  \bibfield  {author} {\bibinfo {author} {\bibfnamefont {W.}~\bibnamefont
  {Wu}}, \bibinfo {author} {\bibfnamefont {J.}~\bibnamefont {Cheng}}, \bibinfo
  {author} {\bibfnamefont {K.}~\bibnamefont {Matsubayashi}}, \bibinfo {author}
  {\bibfnamefont {P.}~\bibnamefont {Kong}}, \bibinfo {author} {\bibfnamefont
  {F.}~\bibnamefont {Lin}}, \bibinfo {author} {\bibfnamefont {C.}~\bibnamefont
  {Jin}}, \bibinfo {author} {\bibfnamefont {N.}~\bibnamefont {Wang}}, \bibinfo
  {author} {\bibfnamefont {Y.}~\bibnamefont {Uwatoko}}, \ and\ \bibinfo
  {author} {\bibfnamefont {J.}~\bibnamefont {Luo}},\ }\href
  {http://dx.doi.org/10.1038/ncomms6508} {\bibfield  {journal} {\bibinfo
  {journal} {Nature Communications}\ }\textbf {\bibinfo {volume} {5}},\
  \bibinfo {pages} {5508} (\bibinfo {year} {2014})}\BibitemShut {NoStop}%
\bibitem [{\citenamefont {Cheng}\ \emph {et~al.}(2015)\citenamefont {Cheng},
  \citenamefont {Matsubayashi}, \citenamefont {Wu}, \citenamefont {Sun},
  \citenamefont {Lin}, \citenamefont {Luo},\ and\ \citenamefont
  {Uwatoko}}]{Cheng2015}%
  \BibitemOpen
  \bibfield  {author} {\bibinfo {author} {\bibfnamefont {J.-G.}\ \bibnamefont
  {Cheng}}, \bibinfo {author} {\bibfnamefont {K.}~\bibnamefont {Matsubayashi}},
  \bibinfo {author} {\bibfnamefont {W.}~\bibnamefont {Wu}}, \bibinfo {author}
  {\bibfnamefont {J.~P.}\ \bibnamefont {Sun}}, \bibinfo {author} {\bibfnamefont
  {F.~K.}\ \bibnamefont {Lin}}, \bibinfo {author} {\bibfnamefont {J.~L.}\
  \bibnamefont {Luo}}, \ and\ \bibinfo {author} {\bibfnamefont
  {Y.}~\bibnamefont {Uwatoko}},\ }\href {\doibase
  10.1103/PhysRevLett.114.117001} {\bibfield  {journal} {\bibinfo  {journal}
  {Phys. Rev. Lett.}\ }\textbf {\bibinfo {volume} {114}},\ \bibinfo {pages}
  {117001} (\bibinfo {year} {2015})}\BibitemShut {NoStop}%
\bibitem [{\citenamefont {Matsuda}\ \emph {et~al.}(2018)\citenamefont
  {Matsuda}, \citenamefont {Lin}, \citenamefont {Yu}, \citenamefont {Cheng},
  \citenamefont {Wu}, \citenamefont {Sun}, \citenamefont {Zhang}, \citenamefont
  {Sun}, \citenamefont {Matsubayashi}, \citenamefont {Miyake}, \citenamefont
  {Kato}, \citenamefont {Yan}, \citenamefont {Stone}, \citenamefont {Si},
  \citenamefont {Luo},\ and\ \citenamefont {Uwatoko}}]{JGCheng2018}%
  \BibitemOpen
  \bibfield  {author} {\bibinfo {author} {\bibfnamefont {M.}~\bibnamefont
  {Matsuda}}, \bibinfo {author} {\bibfnamefont {F.~K.}\ \bibnamefont {Lin}},
  \bibinfo {author} {\bibfnamefont {R.}~\bibnamefont {Yu}}, \bibinfo {author}
  {\bibfnamefont {J.-G.}\ \bibnamefont {Cheng}}, \bibinfo {author}
  {\bibfnamefont {W.}~\bibnamefont {Wu}}, \bibinfo {author} {\bibfnamefont
  {J.~P.}\ \bibnamefont {Sun}}, \bibinfo {author} {\bibfnamefont {J.~H.}\
  \bibnamefont {Zhang}}, \bibinfo {author} {\bibfnamefont {P.~J.}\ \bibnamefont
  {Sun}}, \bibinfo {author} {\bibfnamefont {K.}~\bibnamefont {Matsubayashi}},
  \bibinfo {author} {\bibfnamefont {T.}~\bibnamefont {Miyake}}, \bibinfo
  {author} {\bibfnamefont {T.}~\bibnamefont {Kato}}, \bibinfo {author}
  {\bibfnamefont {J.-Q.}\ \bibnamefont {Yan}}, \bibinfo {author} {\bibfnamefont
  {M.~B.}\ \bibnamefont {Stone}}, \bibinfo {author} {\bibfnamefont
  {Q.}~\bibnamefont {Si}}, \bibinfo {author} {\bibfnamefont {J.~L.}\
  \bibnamefont {Luo}}, \ and\ \bibinfo {author} {\bibfnamefont
  {Y.}~\bibnamefont {Uwatoko}},\ }\href {\doibase 10.1103/PhysRevX.8.031017}
  {\bibfield  {journal} {\bibinfo  {journal} {Phys. Rev. X}\ }\textbf {\bibinfo
  {volume} {8}},\ \bibinfo {pages} {031017} (\bibinfo {year}
  {2018})}\BibitemShut {NoStop}%
\bibitem [{\citenamefont {Cheng}\ and\ \citenamefont {Luo}(2017)}]{Cheng2017}%
  \BibitemOpen
  \bibfield  {author} {\bibinfo {author} {\bibfnamefont {J.}~\bibnamefont
  {Cheng}}\ and\ \bibinfo {author} {\bibfnamefont {J.}~\bibnamefont {Luo}},\
  }\href {http://stacks.iop.org/0953-8984/29/i=38/a=383003} {\bibfield
  {journal} {\bibinfo  {journal} {Journal of Physics: Condensed Matter}\
  }\textbf {\bibinfo {volume} {29}},\ \bibinfo {pages} {383003} (\bibinfo
  {year} {2017})}\BibitemShut {NoStop}%
\bibitem [{\citenamefont {Abanov}\ \emph {et~al.}(2003)\citenamefont {Abanov},
  \citenamefont {Chubukov},\ and\ \citenamefont {Schmalian}}]{Abanov2003}%
  \BibitemOpen
  \bibfield  {author} {\bibinfo {author} {\bibfnamefont {A.}~\bibnamefont
  {Abanov}}, \bibinfo {author} {\bibfnamefont {A.~V.}\ \bibnamefont
  {Chubukov}}, \ and\ \bibinfo {author} {\bibfnamefont {J.}~\bibnamefont
  {Schmalian}},\ }\href {\doibase 10.1080/0001873021000057123} {\bibfield
  {journal} {\bibinfo  {journal} {Advances in Physics}\ }\textbf {\bibinfo
  {volume} {52}},\ \bibinfo {pages} {119} (\bibinfo {year} {2003})}\BibitemShut
  {NoStop}%
\bibitem [{\citenamefont {Abanov}\ and\ \citenamefont
  {Chubukov}(2004)}]{Abanov2004}%
  \BibitemOpen
  \bibfield  {author} {\bibinfo {author} {\bibfnamefont {A.}~\bibnamefont
  {Abanov}}\ and\ \bibinfo {author} {\bibfnamefont {A.}~\bibnamefont
  {Chubukov}},\ }\href {\doibase 10.1103/PhysRevLett.93.255702} {\bibfield
  {journal} {\bibinfo  {journal} {Phys. Rev. Lett.}\ }\textbf {\bibinfo
  {volume} {93}},\ \bibinfo {pages} {255702} (\bibinfo {year}
  {2004})}\BibitemShut {NoStop}%
\bibitem [{\citenamefont {Metlitski}\ and\ \citenamefont
  {Sachdev}(2010{\natexlab{a}})}]{Metlitski2010a}%
  \BibitemOpen
  \bibfield  {author} {\bibinfo {author} {\bibfnamefont {M.~A.}\ \bibnamefont
  {Metlitski}}\ and\ \bibinfo {author} {\bibfnamefont {S.}~\bibnamefont
  {Sachdev}},\ }\href {\doibase 10.1103/PhysRevB.82.075127} {\bibfield
  {journal} {\bibinfo  {journal} {Phys. Rev. B}\ }\textbf {\bibinfo {volume}
  {82}},\ \bibinfo {pages} {075127} (\bibinfo {year}
  {2010}{\natexlab{a}})}\BibitemShut {NoStop}%
\bibitem [{\citenamefont {Metlitski}\ and\ \citenamefont
  {Sachdev}(2010{\natexlab{b}})}]{Metlitski2010b}%
  \BibitemOpen
  \bibfield  {author} {\bibinfo {author} {\bibfnamefont {M.~A.}\ \bibnamefont
  {Metlitski}}\ and\ \bibinfo {author} {\bibfnamefont {S.}~\bibnamefont
  {Sachdev}},\ }\href {\doibase 10.1103/PhysRevB.82.075128} {\bibfield
  {journal} {\bibinfo  {journal} {Phys. Rev. B}\ }\textbf {\bibinfo {volume}
  {82}},\ \bibinfo {pages} {075128} (\bibinfo {year}
  {2010}{\natexlab{b}})}\BibitemShut {NoStop}%
\bibitem [{\citenamefont {Sur}\ and\ \citenamefont {Lee}(2016)}]{Sur2016}%
  \BibitemOpen
  \bibfield  {author} {\bibinfo {author} {\bibfnamefont {S.}~\bibnamefont
  {Sur}}\ and\ \bibinfo {author} {\bibfnamefont {S.-S.}\ \bibnamefont {Lee}},\
  }\href {\doibase 10.1103/PhysRevB.94.195135} {\bibfield  {journal} {\bibinfo
  {journal} {Phys. Rev. B}\ }\textbf {\bibinfo {volume} {94}},\ \bibinfo
  {pages} {195135} (\bibinfo {year} {2016})}\BibitemShut {NoStop}%
\bibitem [{\citenamefont {Schlief}\ \emph {et~al.}(2017)\citenamefont
  {Schlief}, \citenamefont {Lunts},\ and\ \citenamefont {Lee}}]{Schlief2017}%
  \BibitemOpen
  \bibfield  {author} {\bibinfo {author} {\bibfnamefont {A.}~\bibnamefont
  {Schlief}}, \bibinfo {author} {\bibfnamefont {P.}~\bibnamefont {Lunts}}, \
  and\ \bibinfo {author} {\bibfnamefont {S.-S.}\ \bibnamefont {Lee}},\ }\href
  {\doibase 10.1103/PhysRevX.7.021010} {\bibfield  {journal} {\bibinfo
  {journal} {Phys. Rev. X}\ }\textbf {\bibinfo {volume} {7}},\ \bibinfo {pages}
  {021010} (\bibinfo {year} {2017})}\BibitemShut {NoStop}%
\bibitem [{\citenamefont {Lee}(2018)}]{SSLee2018}%
  \BibitemOpen
  \bibfield  {author} {\bibinfo {author} {\bibfnamefont {S.-S.}\ \bibnamefont
  {Lee}},\ }\href {\doibase 10.1146/annurev-conmatphys-031016-025531}
  {\bibfield  {journal} {\bibinfo  {journal} {Annual Review of Condensed Matter
  Physics}\ }\textbf {\bibinfo {volume} {9}},\ \bibinfo {pages} {227} (\bibinfo
  {year} {2018})},\ \Eprint
  {http://arxiv.org/abs/https://doi.org/10.1146/annurev-conmatphys-031016-025531}
  {https://doi.org/10.1146/annurev-conmatphys-031016-025531} \BibitemShut
  {NoStop}%
\bibitem [{\citenamefont {{Schlief}}\ \emph {et~al.}(2018)\citenamefont
  {{Schlief}}, \citenamefont {{Lunts}},\ and\ \citenamefont
  {{Lee}}}]{Schlief2018}%
  \BibitemOpen
  \bibfield  {author} {\bibinfo {author} {\bibfnamefont {A.}~\bibnamefont
  {{Schlief}}}, \bibinfo {author} {\bibfnamefont {P.}~\bibnamefont {{Lunts}}},
  \ and\ \bibinfo {author} {\bibfnamefont {S.-S.}\ \bibnamefont {{Lee}}},\
  }\href@noop {} {\bibfield  {journal} {\bibinfo  {journal} {ArXiv e-prints}\ }
  (\bibinfo {year} {2018})},\ \Eprint {http://arxiv.org/abs/1805.05252}
  {arXiv:1805.05252 [cond-mat.str-el]} \BibitemShut {NoStop}%
\bibitem [{\citenamefont {Chubukov}(2018)}]{Chubukov2018JC}%
  \BibitemOpen
  \bibfield  {author} {\bibinfo {author} {\bibfnamefont {A.}~\bibnamefont
  {Chubukov}},\ }\href {https://www.condmatjclub.org/?p=3482} {\bibfield
  {journal} {\bibinfo  {journal} {Journal Club for Condensed Matter Physics}\
  }\textbf {\bibinfo {volume} {JCCM}},\ \bibinfo {pages} {201802} (\bibinfo
  {year} {2018})}\BibitemShut {NoStop}%
\bibitem [{\citenamefont {{Xu}}\ \emph {et~al.}(2019)\citenamefont {{Xu}},
  \citenamefont {{Liu}}, \citenamefont {{Pan}}, \citenamefont {{Qi}},
  \citenamefont {{Sun}},\ and\ \citenamefont {{Meng}}}]{XiaoYanXu2019_review}%
  \BibitemOpen
  \bibfield  {author} {\bibinfo {author} {\bibfnamefont {X.~Y.}\ \bibnamefont
  {{Xu}}}, \bibinfo {author} {\bibfnamefont {Z.~H.}\ \bibnamefont {{Liu}}},
  \bibinfo {author} {\bibfnamefont {G.}~\bibnamefont {{Pan}}}, \bibinfo
  {author} {\bibfnamefont {Y.}~\bibnamefont {{Qi}}}, \bibinfo {author}
  {\bibfnamefont {K.}~\bibnamefont {{Sun}}}, \ and\ \bibinfo {author}
  {\bibfnamefont {Z.~Y.}\ \bibnamefont {{Meng}}},\ }\href@noop {} {\bibfield
  {journal} {\bibinfo  {journal} {arXiv e-prints}\ ,\ \bibinfo {eid}
  {arXiv:1904.07355}} (\bibinfo {year} {2019})},\ \Eprint
  {http://arxiv.org/abs/1904.07355} {arXiv:1904.07355 [cond-mat.str-el]}
  \BibitemShut {NoStop}%
\bibitem [{\citenamefont {Schattner}\ \emph
  {et~al.}(2016{\natexlab{a}})\citenamefont {Schattner}, \citenamefont
  {Lederer}, \citenamefont {Kivelson},\ and\ \citenamefont
  {Berg}}]{Schattner2015a}%
  \BibitemOpen
  \bibfield  {author} {\bibinfo {author} {\bibfnamefont {Y.}~\bibnamefont
  {Schattner}}, \bibinfo {author} {\bibfnamefont {S.}~\bibnamefont {Lederer}},
  \bibinfo {author} {\bibfnamefont {S.~A.}\ \bibnamefont {Kivelson}}, \ and\
  \bibinfo {author} {\bibfnamefont {E.}~\bibnamefont {Berg}},\ }\href {\doibase
  10.1103/PhysRevX.6.031028} {\bibfield  {journal} {\bibinfo  {journal} {Phys.
  Rev. X}\ }\textbf {\bibinfo {volume} {6}},\ \bibinfo {pages} {031028}
  (\bibinfo {year} {2016}{\natexlab{a}})}\BibitemShut {NoStop}%
\bibitem [{\citenamefont {Lederer}\ \emph {et~al.}(2017)\citenamefont
  {Lederer}, \citenamefont {Schattner}, \citenamefont {Berg},\ and\
  \citenamefont {Kivelson}}]{Lederer2016}%
  \BibitemOpen
  \bibfield  {author} {\bibinfo {author} {\bibfnamefont {S.}~\bibnamefont
  {Lederer}}, \bibinfo {author} {\bibfnamefont {Y.}~\bibnamefont {Schattner}},
  \bibinfo {author} {\bibfnamefont {E.}~\bibnamefont {Berg}}, \ and\ \bibinfo
  {author} {\bibfnamefont {S.~A.}\ \bibnamefont {Kivelson}},\ }\href {\doibase
  10.1073/pnas.1620651114} {\bibfield  {journal} {\bibinfo  {journal}
  {Proceedings of the National Academy of Sciences}\ } (\bibinfo {year}
  {2017}),\ 10.1073/pnas.1620651114}\BibitemShut {NoStop}%
\bibitem [{\citenamefont {{Li}}\ \emph {et~al.}(2015)\citenamefont {{Li}},
  \citenamefont {{Wang}}, \citenamefont {{Yao}},\ and\ \citenamefont
  {{Lee}}}]{ZXLi2015}%
  \BibitemOpen
  \bibfield  {author} {\bibinfo {author} {\bibfnamefont {Z.-X.}\ \bibnamefont
  {{Li}}}, \bibinfo {author} {\bibfnamefont {F.}~\bibnamefont {{Wang}}},
  \bibinfo {author} {\bibfnamefont {H.}~\bibnamefont {{Yao}}}, \ and\ \bibinfo
  {author} {\bibfnamefont {D.-H.}\ \bibnamefont {{Lee}}},\ }\href@noop {}
  {\bibfield  {journal} {\bibinfo  {journal} {ArXiv e-prints}\ } (\bibinfo
  {year} {2015})},\ \Eprint {http://arxiv.org/abs/1512.04541} {arXiv:1512.04541
  [cond-mat.supr-con]} \BibitemShut {NoStop}%
\bibitem [{\citenamefont {Berg}\ \emph {et~al.}(2012)\citenamefont {Berg},
  \citenamefont {Metlitski},\ and\ \citenamefont {Sachdev}}]{Berg12}%
  \BibitemOpen
  \bibfield  {author} {\bibinfo {author} {\bibfnamefont {E.}~\bibnamefont
  {Berg}}, \bibinfo {author} {\bibfnamefont {M.~A.}\ \bibnamefont {Metlitski}},
  \ and\ \bibinfo {author} {\bibfnamefont {S.}~\bibnamefont {Sachdev}},\ }\href
  {\doibase 10.1126/science.1227769} {\bibfield  {journal} {\bibinfo  {journal}
  {Science}\ }\textbf {\bibinfo {volume} {338}},\ \bibinfo {pages} {1606}
  (\bibinfo {year} {2012})}\BibitemShut {NoStop}%
\bibitem [{\citenamefont {Li}\ \emph {et~al.}(2016)\citenamefont {Li},
  \citenamefont {Wang}, \citenamefont {Yao},\ and\ \citenamefont
  {Lee}}]{ZXLi2016}%
  \BibitemOpen
  \bibfield  {author} {\bibinfo {author} {\bibfnamefont {Z.-X.}\ \bibnamefont
  {Li}}, \bibinfo {author} {\bibfnamefont {F.}~\bibnamefont {Wang}}, \bibinfo
  {author} {\bibfnamefont {H.}~\bibnamefont {Yao}}, \ and\ \bibinfo {author}
  {\bibfnamefont {D.-H.}\ \bibnamefont {Lee}},\ }\href {\doibase
  http://dx.doi.org/10.1007/s11434-016-1087-x} {\bibfield  {journal} {\bibinfo
  {journal} {Science Bulletin}\ }\textbf {\bibinfo {volume} {61}},\ \bibinfo
  {pages} {925 } (\bibinfo {year} {2016})}\BibitemShut {NoStop}%
\bibitem [{\citenamefont {Schattner}\ \emph
  {et~al.}(2016{\natexlab{b}})\citenamefont {Schattner}, \citenamefont
  {Gerlach}, \citenamefont {Trebst},\ and\ \citenamefont
  {Berg}}]{Schattner2015b}%
  \BibitemOpen
  \bibfield  {author} {\bibinfo {author} {\bibfnamefont {Y.}~\bibnamefont
  {Schattner}}, \bibinfo {author} {\bibfnamefont {M.~H.}\ \bibnamefont
  {Gerlach}}, \bibinfo {author} {\bibfnamefont {S.}~\bibnamefont {Trebst}}, \
  and\ \bibinfo {author} {\bibfnamefont {E.}~\bibnamefont {Berg}},\ }\href
  {\doibase 10.1103/PhysRevLett.117.097002} {\bibfield  {journal} {\bibinfo
  {journal} {Phys. Rev. Lett.}\ }\textbf {\bibinfo {volume} {117}},\ \bibinfo
  {pages} {097002} (\bibinfo {year} {2016}{\natexlab{b}})}\BibitemShut
  {NoStop}%
\bibitem [{\citenamefont {Gerlach}\ \emph {et~al.}(2017)\citenamefont
  {Gerlach}, \citenamefont {Schattner}, \citenamefont {Berg},\ and\
  \citenamefont {Trebst}}]{Gerlach2017}%
  \BibitemOpen
  \bibfield  {author} {\bibinfo {author} {\bibfnamefont {M.~H.}\ \bibnamefont
  {Gerlach}}, \bibinfo {author} {\bibfnamefont {Y.}~\bibnamefont {Schattner}},
  \bibinfo {author} {\bibfnamefont {E.}~\bibnamefont {Berg}}, \ and\ \bibinfo
  {author} {\bibfnamefont {S.}~\bibnamefont {Trebst}},\ }\href {\doibase
  10.1103/PhysRevB.95.035124} {\bibfield  {journal} {\bibinfo  {journal} {Phys.
  Rev. B}\ }\textbf {\bibinfo {volume} {95}},\ \bibinfo {pages} {035124}
  (\bibinfo {year} {2017})}\BibitemShut {NoStop}%
\bibitem [{\citenamefont {Liu}\ \emph {et~al.}(2018)\citenamefont {Liu},
  \citenamefont {Xu}, \citenamefont {Qi}, \citenamefont {Sun},\ and\
  \citenamefont {Meng}}]{ZHLiu2017}%
  \BibitemOpen
  \bibfield  {author} {\bibinfo {author} {\bibfnamefont {Z.~H.}\ \bibnamefont
  {Liu}}, \bibinfo {author} {\bibfnamefont {X.~Y.}\ \bibnamefont {Xu}},
  \bibinfo {author} {\bibfnamefont {Y.}~\bibnamefont {Qi}}, \bibinfo {author}
  {\bibfnamefont {K.}~\bibnamefont {Sun}}, \ and\ \bibinfo {author}
  {\bibfnamefont {Z.~Y.}\ \bibnamefont {Meng}},\ }\href {\doibase
  10.1103/PhysRevB.98.045116} {\bibfield  {journal} {\bibinfo  {journal} {Phys.
  Rev. B}\ }\textbf {\bibinfo {volume} {98}},\ \bibinfo {pages} {045116}
  (\bibinfo {year} {2018})}\BibitemShut {NoStop}%
\bibitem [{\citenamefont {Liu}\ \emph {et~al.}(2019)\citenamefont {Liu},
  \citenamefont {Xu}, \citenamefont {Qi}, \citenamefont {Sun},\ and\
  \citenamefont {Meng}}]{ZHLiu2018}%
  \BibitemOpen
  \bibfield  {author} {\bibinfo {author} {\bibfnamefont {Z.~H.}\ \bibnamefont
  {Liu}}, \bibinfo {author} {\bibfnamefont {X.~Y.}\ \bibnamefont {Xu}},
  \bibinfo {author} {\bibfnamefont {Y.}~\bibnamefont {Qi}}, \bibinfo {author}
  {\bibfnamefont {K.}~\bibnamefont {Sun}}, \ and\ \bibinfo {author}
  {\bibfnamefont {Z.~Y.}\ \bibnamefont {Meng}},\ }\href {\doibase
  10.1103/PhysRevB.99.085114} {\bibfield  {journal} {\bibinfo  {journal} {Phys.
  Rev. B}\ }\textbf {\bibinfo {volume} {99}},\ \bibinfo {pages} {085114}
  (\bibinfo {year} {2019})}\BibitemShut {NoStop}%
\bibitem [{\citenamefont {Xu}\ \emph {et~al.}(2017{\natexlab{b}})\citenamefont
  {Xu}, \citenamefont {Beach}, \citenamefont {Sun}, \citenamefont {Assaad},\
  and\ \citenamefont {Meng}}]{Xu2016a}%
  \BibitemOpen
  \bibfield  {author} {\bibinfo {author} {\bibfnamefont {X.~Y.}\ \bibnamefont
  {Xu}}, \bibinfo {author} {\bibfnamefont {K.~S.~D.}\ \bibnamefont {Beach}},
  \bibinfo {author} {\bibfnamefont {K.}~\bibnamefont {Sun}}, \bibinfo {author}
  {\bibfnamefont {F.~F.}\ \bibnamefont {Assaad}}, \ and\ \bibinfo {author}
  {\bibfnamefont {Z.~Y.}\ \bibnamefont {Meng}},\ }\href {\doibase
  10.1103/PhysRevB.95.085110} {\bibfield  {journal} {\bibinfo  {journal} {Phys.
  Rev. B}\ }\textbf {\bibinfo {volume} {95}},\ \bibinfo {pages} {085110}
  (\bibinfo {year} {2017}{\natexlab{b}})}\BibitemShut {NoStop}%
\bibitem [{\citenamefont {Assaad}\ and\ \citenamefont
  {Grover}(2016)}]{Assaad2016}%
  \BibitemOpen
  \bibfield  {author} {\bibinfo {author} {\bibfnamefont {F.~F.}\ \bibnamefont
  {Assaad}}\ and\ \bibinfo {author} {\bibfnamefont {T.}~\bibnamefont
  {Grover}},\ }\href {\doibase 10.1103/PhysRevX.6.041049} {\bibfield  {journal}
  {\bibinfo  {journal} {Phys. Rev. X}\ }\textbf {\bibinfo {volume} {6}},\
  \bibinfo {pages} {041049} (\bibinfo {year} {2016})}\BibitemShut {NoStop}%
\bibitem [{\citenamefont {Gazit}\ \emph {et~al.}(2017)\citenamefont {Gazit},
  \citenamefont {Randeria},\ and\ \citenamefont {Vishwanath}}]{Gazit2016}%
  \BibitemOpen
  \bibfield  {author} {\bibinfo {author} {\bibfnamefont {S.}~\bibnamefont
  {Gazit}}, \bibinfo {author} {\bibfnamefont {M.}~\bibnamefont {Randeria}}, \
  and\ \bibinfo {author} {\bibfnamefont {A.}~\bibnamefont {Vishwanath}},\
  }\href {http://dx.doi.org/10.1038/nphys4028} {\bibfield  {journal} {\bibinfo
  {journal} {Nat Phys}\ }\textbf {\bibinfo {volume} {advance online
  publication}} (\bibinfo {year} {2017})}\BibitemShut {NoStop}%
\bibitem [{\citenamefont {He}\ \emph {et~al.}(2018)\citenamefont {He},
  \citenamefont {Xu}, \citenamefont {Sun}, \citenamefont {Assaad},
  \citenamefont {Meng},\ and\ \citenamefont {Lu}}]{He2018}%
  \BibitemOpen
  \bibfield  {author} {\bibinfo {author} {\bibfnamefont {Y.-Y.}\ \bibnamefont
  {He}}, \bibinfo {author} {\bibfnamefont {X.~Y.}\ \bibnamefont {Xu}}, \bibinfo
  {author} {\bibfnamefont {K.}~\bibnamefont {Sun}}, \bibinfo {author}
  {\bibfnamefont {F.~F.}\ \bibnamefont {Assaad}}, \bibinfo {author}
  {\bibfnamefont {Z.~Y.}\ \bibnamefont {Meng}}, \ and\ \bibinfo {author}
  {\bibfnamefont {Z.-Y.}\ \bibnamefont {Lu}},\ }\href {\doibase
  10.1103/PhysRevB.97.081110} {\bibfield  {journal} {\bibinfo  {journal} {Phys.
  Rev. B}\ }\textbf {\bibinfo {volume} {97}},\ \bibinfo {pages} {081110}
  (\bibinfo {year} {2018})}\BibitemShut {NoStop}%
\bibitem [{\citenamefont {Xu}\ \emph {et~al.}(2019)\citenamefont {Xu},
  \citenamefont {Qi}, \citenamefont {Zhang}, \citenamefont {Assaad},
  \citenamefont {Xu},\ and\ \citenamefont {Meng}}]{XiaoYanXu2019_PRX}%
  \BibitemOpen
  \bibfield  {author} {\bibinfo {author} {\bibfnamefont {X.~Y.}\ \bibnamefont
  {Xu}}, \bibinfo {author} {\bibfnamefont {Y.}~\bibnamefont {Qi}}, \bibinfo
  {author} {\bibfnamefont {L.}~\bibnamefont {Zhang}}, \bibinfo {author}
  {\bibfnamefont {F.~F.}\ \bibnamefont {Assaad}}, \bibinfo {author}
  {\bibfnamefont {C.}~\bibnamefont {Xu}}, \ and\ \bibinfo {author}
  {\bibfnamefont {Z.~Y.}\ \bibnamefont {Meng}},\ }\href {\doibase
  10.1103/PhysRevX.9.021022} {\bibfield  {journal} {\bibinfo  {journal} {Phys.
  Rev. X}\ }\textbf {\bibinfo {volume} {9}},\ \bibinfo {pages} {021022}
  (\bibinfo {year} {2019})}\BibitemShut {NoStop}%
\bibitem [{\citenamefont {{Chen}}\ \emph {et~al.}(2019)\citenamefont {{Chen}},
  \citenamefont {{Xu}}, \citenamefont {{Qi}},\ and\ \citenamefont
  {{Meng}}}]{ChuangChen2019OM}%
  \BibitemOpen
  \bibfield  {author} {\bibinfo {author} {\bibfnamefont {C.}~\bibnamefont
  {{Chen}}}, \bibinfo {author} {\bibfnamefont {X.~Y.}\ \bibnamefont {{Xu}}},
  \bibinfo {author} {\bibfnamefont {Y.}~\bibnamefont {{Qi}}}, \ and\ \bibinfo
  {author} {\bibfnamefont {Z.~Y.}\ \bibnamefont {{Meng}}},\ }\href@noop {}
  {\bibfield  {journal} {\bibinfo  {journal} {arXiv e-prints}\ ,\ \bibinfo
  {eid} {arXiv:1904.12872}} (\bibinfo {year} {2019})},\ \Eprint
  {http://arxiv.org/abs/1904.12872} {arXiv:1904.12872 [cond-mat.str-el]}
  \BibitemShut {NoStop}%
\bibitem [{\citenamefont {Liu}\ \emph {et~al.}(2017{\natexlab{a}})\citenamefont
  {Liu}, \citenamefont {Qi}, \citenamefont {Meng},\ and\ \citenamefont
  {Fu}}]{liu2016self}%
  \BibitemOpen
  \bibfield  {author} {\bibinfo {author} {\bibfnamefont {J.}~\bibnamefont
  {Liu}}, \bibinfo {author} {\bibfnamefont {Y.}~\bibnamefont {Qi}}, \bibinfo
  {author} {\bibfnamefont {Z.~Y.}\ \bibnamefont {Meng}}, \ and\ \bibinfo
  {author} {\bibfnamefont {L.}~\bibnamefont {Fu}},\ }\href {\doibase
  10.1103/PhysRevB.95.041101} {\bibfield  {journal} {\bibinfo  {journal} {Phys.
  Rev. B}\ }\textbf {\bibinfo {volume} {95}},\ \bibinfo {pages} {041101}
  (\bibinfo {year} {2017}{\natexlab{a}})}\BibitemShut {NoStop}%
\bibitem [{\citenamefont {Liu}\ \emph {et~al.}(2017{\natexlab{b}})\citenamefont
  {Liu}, \citenamefont {Shen}, \citenamefont {Qi}, \citenamefont {Meng},\ and\
  \citenamefont {Fu}}]{liu2016fermion}%
  \BibitemOpen
  \bibfield  {author} {\bibinfo {author} {\bibfnamefont {J.}~\bibnamefont
  {Liu}}, \bibinfo {author} {\bibfnamefont {H.}~\bibnamefont {Shen}}, \bibinfo
  {author} {\bibfnamefont {Y.}~\bibnamefont {Qi}}, \bibinfo {author}
  {\bibfnamefont {Z.~Y.}\ \bibnamefont {Meng}}, \ and\ \bibinfo {author}
  {\bibfnamefont {L.}~\bibnamefont {Fu}},\ }\href {\doibase
  10.1103/PhysRevB.95.241104} {\bibfield  {journal} {\bibinfo  {journal} {Phys.
  Rev. B}\ }\textbf {\bibinfo {volume} {95}},\ \bibinfo {pages} {241104}
  (\bibinfo {year} {2017}{\natexlab{b}})}\BibitemShut {NoStop}%
\bibitem [{\citenamefont {Xu}\ \emph {et~al.}(2017{\natexlab{c}})\citenamefont
  {Xu}, \citenamefont {Qi}, \citenamefont {Liu}, \citenamefont {Fu},\ and\
  \citenamefont {Meng}}]{Xu2016self}%
  \BibitemOpen
  \bibfield  {author} {\bibinfo {author} {\bibfnamefont {X.~Y.}\ \bibnamefont
  {Xu}}, \bibinfo {author} {\bibfnamefont {Y.}~\bibnamefont {Qi}}, \bibinfo
  {author} {\bibfnamefont {J.}~\bibnamefont {Liu}}, \bibinfo {author}
  {\bibfnamefont {L.}~\bibnamefont {Fu}}, \ and\ \bibinfo {author}
  {\bibfnamefont {Z.~Y.}\ \bibnamefont {Meng}},\ }\href {\doibase
  10.1103/PhysRevB.96.041119} {\bibfield  {journal} {\bibinfo  {journal} {Phys.
  Rev. B}\ }\textbf {\bibinfo {volume} {96}},\ \bibinfo {pages} {041119}
  (\bibinfo {year} {2017}{\natexlab{c}})}\BibitemShut {NoStop}%
\bibitem [{\citenamefont {Nagai}\ \emph {et~al.}(2017)\citenamefont {Nagai},
  \citenamefont {Shen}, \citenamefont {Qi}, \citenamefont {Liu},\ and\
  \citenamefont {Fu}}]{Nagai2017}%
  \BibitemOpen
  \bibfield  {author} {\bibinfo {author} {\bibfnamefont {Y.}~\bibnamefont
  {Nagai}}, \bibinfo {author} {\bibfnamefont {H.}~\bibnamefont {Shen}},
  \bibinfo {author} {\bibfnamefont {Y.}~\bibnamefont {Qi}}, \bibinfo {author}
  {\bibfnamefont {J.}~\bibnamefont {Liu}}, \ and\ \bibinfo {author}
  {\bibfnamefont {L.}~\bibnamefont {Fu}},\ }\href {\doibase
  10.1103/PhysRevB.96.161102} {\bibfield  {journal} {\bibinfo  {journal} {Phys.
  Rev. B}\ }\textbf {\bibinfo {volume} {96}},\ \bibinfo {pages} {161102}
  (\bibinfo {year} {2017})}\BibitemShut {NoStop}%
\bibitem [{\citenamefont {Shen}\ \emph {et~al.}(2018)\citenamefont {Shen},
  \citenamefont {Liu},\ and\ \citenamefont {Fu}}]{HTShen2018}%
  \BibitemOpen
  \bibfield  {author} {\bibinfo {author} {\bibfnamefont {H.}~\bibnamefont
  {Shen}}, \bibinfo {author} {\bibfnamefont {J.}~\bibnamefont {Liu}}, \ and\
  \bibinfo {author} {\bibfnamefont {L.}~\bibnamefont {Fu}},\ }\href {\doibase
  10.1103/PhysRevB.97.205140} {\bibfield  {journal} {\bibinfo  {journal} {Phys.
  Rev. B}\ }\textbf {\bibinfo {volume} {97}},\ \bibinfo {pages} {205140}
  (\bibinfo {year} {2018})}\BibitemShut {NoStop}%
\bibitem [{\citenamefont {Chen}\ \emph {et~al.}(2018)\citenamefont {Chen},
  \citenamefont {Xu}, \citenamefont {Liu}, \citenamefont {Batrouni},
  \citenamefont {Scalettar},\ and\ \citenamefont {Meng}}]{ChenChuang2018}%
  \BibitemOpen
  \bibfield  {author} {\bibinfo {author} {\bibfnamefont {C.}~\bibnamefont
  {Chen}}, \bibinfo {author} {\bibfnamefont {X.~Y.}\ \bibnamefont {Xu}},
  \bibinfo {author} {\bibfnamefont {J.}~\bibnamefont {Liu}}, \bibinfo {author}
  {\bibfnamefont {G.}~\bibnamefont {Batrouni}}, \bibinfo {author}
  {\bibfnamefont {R.}~\bibnamefont {Scalettar}}, \ and\ \bibinfo {author}
  {\bibfnamefont {Z.~Y.}\ \bibnamefont {Meng}},\ }\href {\doibase
  10.1103/PhysRevB.98.041102} {\bibfield  {journal} {\bibinfo  {journal} {Phys.
  Rev. B}\ }\textbf {\bibinfo {volume} {98}},\ \bibinfo {pages} {041102}
  (\bibinfo {year} {2018})}\BibitemShut {NoStop}%
\bibitem [{\citenamefont {Chen}\ \emph {et~al.}(2019)\citenamefont {Chen},
  \citenamefont {Xu}, \citenamefont {Meng},\ and\ \citenamefont
  {Hohenadler}}]{ChenChuang2018Dirac}%
  \BibitemOpen
  \bibfield  {author} {\bibinfo {author} {\bibfnamefont {C.}~\bibnamefont
  {Chen}}, \bibinfo {author} {\bibfnamefont {X.~Y.}\ \bibnamefont {Xu}},
  \bibinfo {author} {\bibfnamefont {Z.~Y.}\ \bibnamefont {Meng}}, \ and\
  \bibinfo {author} {\bibfnamefont {M.}~\bibnamefont {Hohenadler}},\ }\href
  {\doibase 10.1103/PhysRevLett.122.077601} {\bibfield  {journal} {\bibinfo
  {journal} {Phys. Rev. Lett.}\ }\textbf {\bibinfo {volume} {122}},\ \bibinfo
  {pages} {077601} (\bibinfo {year} {2019})}\BibitemShut {NoStop}%
\bibitem [{\citenamefont {Sun}\ \emph {et~al.}(2008)\citenamefont {Sun},
  \citenamefont {Fregoso}, \citenamefont {Lawler},\ and\ \citenamefont
  {Fradkin}}]{KaiSun2008}%
  \BibitemOpen
  \bibfield  {author} {\bibinfo {author} {\bibfnamefont {K.}~\bibnamefont
  {Sun}}, \bibinfo {author} {\bibfnamefont {B.~M.}\ \bibnamefont {Fregoso}},
  \bibinfo {author} {\bibfnamefont {M.~J.}\ \bibnamefont {Lawler}}, \ and\
  \bibinfo {author} {\bibfnamefont {E.}~\bibnamefont {Fradkin}},\ }\href
  {\doibase 10.1103/PhysRevB.78.085124} {\bibfield  {journal} {\bibinfo
  {journal} {Phys. Rev. B}\ }\textbf {\bibinfo {volume} {78}},\ \bibinfo
  {pages} {085124} (\bibinfo {year} {2008})}\BibitemShut {NoStop}%
\bibitem [{\citenamefont {Berg}\ \emph {et~al.}(2019)\citenamefont {Berg},
  \citenamefont {Lederer}, \citenamefont {Schattner},\ and\ \citenamefont
  {Trebst}}]{Berg2018}%
  \BibitemOpen
  \bibfield  {author} {\bibinfo {author} {\bibfnamefont {E.}~\bibnamefont
  {Berg}}, \bibinfo {author} {\bibfnamefont {S.}~\bibnamefont {Lederer}},
  \bibinfo {author} {\bibfnamefont {Y.}~\bibnamefont {Schattner}}, \ and\
  \bibinfo {author} {\bibfnamefont {S.}~\bibnamefont {Trebst}},\ }\href
  {\doibase 10.1146/annurev-conmatphys-031218-013339} {\bibfield  {journal}
  {\bibinfo  {journal} {Annual Review of Condensed Matter Physics}\ }\textbf
  {\bibinfo {volume} {10}},\ \bibinfo {pages} {63} (\bibinfo {year} {2019})},\
  \Eprint
  {http://arxiv.org/abs/https://doi.org/10.1146/annurev-conmatphys-031218-013339}
  {https://doi.org/10.1146/annurev-conmatphys-031218-013339} \BibitemShut
  {NoStop}%
\bibitem [{\citenamefont {Bl\"ote}\ and\ \citenamefont
  {Deng}(2002)}]{Bloete2002}%
  \BibitemOpen
  \bibfield  {author} {\bibinfo {author} {\bibfnamefont {H.~W.~J.}\
  \bibnamefont {Bl\"ote}}\ and\ \bibinfo {author} {\bibfnamefont
  {Y.}~\bibnamefont {Deng}},\ }\href {\doibase 10.1103/PhysRevE.66.066110}
  {\bibfield  {journal} {\bibinfo  {journal} {Phys. Rev. E}\ }\textbf {\bibinfo
  {volume} {66}},\ \bibinfo {pages} {066110} (\bibinfo {year}
  {2002})}\BibitemShut {NoStop}%
\bibitem [{\citenamefont {Hesselmann}\ and\ \citenamefont
  {Wessel}(2016)}]{Hesselmann2016}%
  \BibitemOpen
  \bibfield  {author} {\bibinfo {author} {\bibfnamefont {S.}~\bibnamefont
  {Hesselmann}}\ and\ \bibinfo {author} {\bibfnamefont {S.}~\bibnamefont
  {Wessel}},\ }\href {\doibase 10.1103/PhysRevB.93.155157} {\bibfield
  {journal} {\bibinfo  {journal} {Phys. Rev. B}\ }\textbf {\bibinfo {volume}
  {93}},\ \bibinfo {pages} {155157} (\bibinfo {year} {2016})}\BibitemShut
  {NoStop}%
\bibitem [{\citenamefont {Chowdhury}\ and\ \citenamefont
  {Sachdev}(2014)}]{Chowdhury2014}%
  \BibitemOpen
  \bibfield  {author} {\bibinfo {author} {\bibfnamefont {D.}~\bibnamefont
  {Chowdhury}}\ and\ \bibinfo {author} {\bibfnamefont {S.}~\bibnamefont
  {Sachdev}},\ }\href {\doibase 10.1103/PhysRevB.90.245136} {\bibfield
  {journal} {\bibinfo  {journal} {Phys. Rev. B}\ }\textbf {\bibinfo {volume}
  {90}},\ \bibinfo {pages} {245136} (\bibinfo {year} {2014})}\BibitemShut
  {NoStop}%
\bibitem [{\citenamefont {Wang}\ \emph {et~al.}(2017)\citenamefont {Wang},
  \citenamefont {Qi}, \citenamefont {Chen},\ and\ \citenamefont
  {Meng}}]{YCWang2017}%
  \BibitemOpen
  \bibfield  {author} {\bibinfo {author} {\bibfnamefont {Y.-C.}\ \bibnamefont
  {Wang}}, \bibinfo {author} {\bibfnamefont {Y.}~\bibnamefont {Qi}}, \bibinfo
  {author} {\bibfnamefont {S.}~\bibnamefont {Chen}}, \ and\ \bibinfo {author}
  {\bibfnamefont {Z.~Y.}\ \bibnamefont {Meng}},\ }\href {\doibase
  10.1103/PhysRevB.96.115160} {\bibfield  {journal} {\bibinfo  {journal} {Phys.
  Rev. B}\ }\textbf {\bibinfo {volume} {96}},\ \bibinfo {pages} {115160}
  (\bibinfo {year} {2017})}\BibitemShut {NoStop}%
\bibitem [{\citenamefont {Wolff}(1989)}]{Wolff1989}%
  \BibitemOpen
  \bibfield  {author} {\bibinfo {author} {\bibfnamefont {U.}~\bibnamefont
  {Wolff}},\ }\href {\doibase 10.1103/PhysRevLett.62.361} {\bibfield  {journal}
  {\bibinfo  {journal} {Phys. Rev. Lett.}\ }\textbf {\bibinfo {volume} {62}},\
  \bibinfo {pages} {361} (\bibinfo {year} {1989})}\BibitemShut {NoStop}%
\bibitem [{\citenamefont {Swendsen}\ and\ \citenamefont
  {Wang}(1987)}]{Swendsen1987}%
  \BibitemOpen
  \bibfield  {author} {\bibinfo {author} {\bibfnamefont {R.~H.}\ \bibnamefont
  {Swendsen}}\ and\ \bibinfo {author} {\bibfnamefont {J.-S.}\ \bibnamefont
  {Wang}},\ }\href {\doibase 10.1103/PhysRevLett.58.86} {\bibfield  {journal}
  {\bibinfo  {journal} {Phys. Rev. Lett.}\ }\textbf {\bibinfo {volume} {58}},\
  \bibinfo {pages} {86} (\bibinfo {year} {1987})}\BibitemShut {NoStop}%
\bibitem [{\citenamefont {Blankenbecler}\ \emph {et~al.}(1981)\citenamefont
  {Blankenbecler}, \citenamefont {Scalapino},\ and\ \citenamefont
  {Sugar}}]{BSS1981}%
  \BibitemOpen
  \bibfield  {author} {\bibinfo {author} {\bibfnamefont {R.}~\bibnamefont
  {Blankenbecler}}, \bibinfo {author} {\bibfnamefont {D.~J.}\ \bibnamefont
  {Scalapino}}, \ and\ \bibinfo {author} {\bibfnamefont {R.~L.}\ \bibnamefont
  {Sugar}},\ }\href {\doibase 10.1103/PhysRevD.24.2278} {\bibfield  {journal}
  {\bibinfo  {journal} {Phys. Rev. D}\ }\textbf {\bibinfo {volume} {24}},\
  \bibinfo {pages} {2278} (\bibinfo {year} {1981})}\BibitemShut {NoStop}%
\bibitem [{\citenamefont {Hirsch}(1983)}]{Hirsch_1983}%
  \BibitemOpen
  \bibfield  {author} {\bibinfo {author} {\bibfnamefont {J.~E.}\ \bibnamefont
  {Hirsch}},\ }\href {\doibase 10.1103/physrevb.28.4059} {\bibfield  {journal}
  {\bibinfo  {journal} {Physical Review B}\ }\textbf {\bibinfo {volume} {28}},\
  \bibinfo {pages} {4059} (\bibinfo {year} {1983})}\BibitemShut {NoStop}%
\bibitem [{\citenamefont {Assaad}\ and\ \citenamefont
  {Evertz}(2008)}]{AssaadEvertz2008}%
  \BibitemOpen
  \bibfield  {author} {\bibinfo {author} {\bibfnamefont {F.}~\bibnamefont
  {Assaad}}\ and\ \bibinfo {author} {\bibfnamefont {H.}~\bibnamefont
  {Evertz}},\ }in\ \href {\doibase 10.1007/978-3-540-74686-7_10} {\emph
  {\bibinfo {booktitle} {Computational Many-Particle Physics}}},\ \bibinfo
  {series} {Lecture Notes in Physics}, Vol.\ \bibinfo {volume} {739},\ \bibinfo
  {editor} {edited by\ \bibinfo {editor} {\bibfnamefont {H.}~\bibnamefont
  {Fehske}}, \bibinfo {editor} {\bibfnamefont {R.}~\bibnamefont {Schneider}}, \
  and\ \bibinfo {editor} {\bibfnamefont {A.}~\bibnamefont {Wei{\ss}e}}}\
  (\bibinfo  {publisher} {Springer Berlin Heidelberg},\ \bibinfo {year}
  {2008})\ pp.\ \bibinfo {pages} {277--356}\BibitemShut {NoStop}%
\bibitem [{\citenamefont {Metlitski}\ and\ \citenamefont
  {Sachdev}(2010{\natexlab{c}})}]{Metlitski2010}%
  \BibitemOpen
  \bibfield  {author} {\bibinfo {author} {\bibfnamefont {M.~A.}\ \bibnamefont
  {Metlitski}}\ and\ \bibinfo {author} {\bibfnamefont {S.}~\bibnamefont
  {Sachdev}},\ }\href {http://stacks.iop.org/1367-2630/12/i=10/a=105007}
  {\bibfield  {journal} {\bibinfo  {journal} {New Journal of Physics}\ }\textbf
  {\bibinfo {volume} {12}},\ \bibinfo {pages} {105007} (\bibinfo {year}
  {2010}{\natexlab{c}})}\BibitemShut {NoStop}%
\end{thebibliography}%

\appendix

\section{THERMAL PHASE TRANSITION}
The thermal phase boundary in Fig. 1(c) of the main text is controlled by the 2D Ising critical exponents $\beta=1/8$ and $\nu=1$. In our model, the AFM Ising field order parameter is given by

\begin{equation}
|m|=\left|\int d\tau\sum_{i}s_{i}^{z}(\tau)e^{i\mathbf{Q}\cdot\mathbf{r}_{i}}\right|,
\label{order parameter}
\end{equation}
where $\mathbf{Q}=(\pi,\pi)$ represented the finite $2\mathbf{Q}=\Gamma$ SDW. Close to the thermal phase boundary, the order parameter $|m|$ are expected to obey the finite size scaling forms

\begin{equation}
m(h,\,L)=L^{-\beta/\nu}f(L^{1/\nu}(h-h_{c})),
\label{eq:datacollapseofm}
\end{equation}
where $f$ is the scaling function. To locate the finite temperature classical critical point, we collect the $m(h,L)$ with different size $L$ and transverse field $h$, and rescale the order parameter as

\begin{equation}
L^{\beta/\nu}m(h,\,L)=f(L^{1/\nu}(h-h_{c})).
\label{eq:rescaleorder}
\end{equation}

\begin{figure}[htp]
	\centering
	\includegraphics[width=0.45\textwidth]{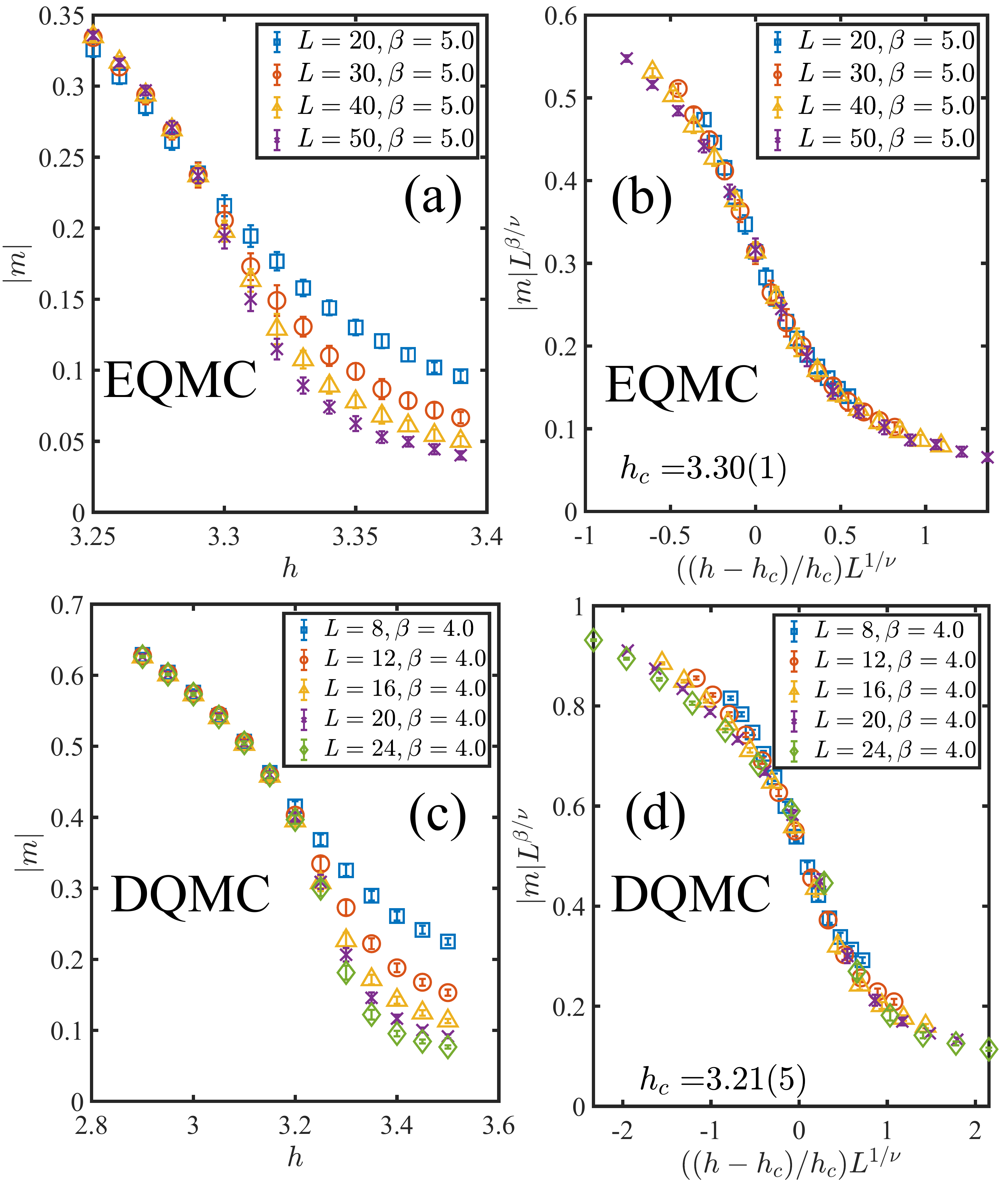}
	\caption{Finite-temperature AFM-to-PM phase transition and the date collapses according to the 2D Ising critical exponents for EQMC simulations (upper panels (a) and (b)) and DQMC simulations (lower planes (c) and (d)).}
	\label{FTtransition}
\end{figure}

In the main text, we show the finite temperature phase boundary obtained by EQMC and DQMC. Fig.~\ref{FTtransition} shows the examples of how these finite temperature phase boundaries are determined. Fig.~\ref{FTtransition} (a) and (b) are for the EQMC results at $\beta=5$ and Fig.~\ref{FTtransition} (c) and (d) are for the DQMC results at $\beta=4$. 

The left panels (a) and (c) plot the original data of the order parameter $|m(h,L)|$ vs $h$. The right panels (b) and (d) shows the data collapse according to Eq.~\eqref{eq:datacollapseofm}. Note that $h_c$ is a free fitting parameter, and the best data collapse gives $h_c=3.30(1)$ at $\beta=5.0$ for EQMC and $h_c=3.21(5)$ at $\beta=4.0$ for DQMC. 

All the other critical field strength $h_c$ at finite temperature phase transitions, given in pure boson model (the 2D transverse field Ising model), DQMC and EQMC simulations, presented in the Fig.1 (c) of the main text, are obtained in the same procedure as shown here.

\section{Comparison of EQMC and DQMC at quantum critical region}
\begin{figure}[htp]
	\centering
	\includegraphics[width=0.45\textwidth]{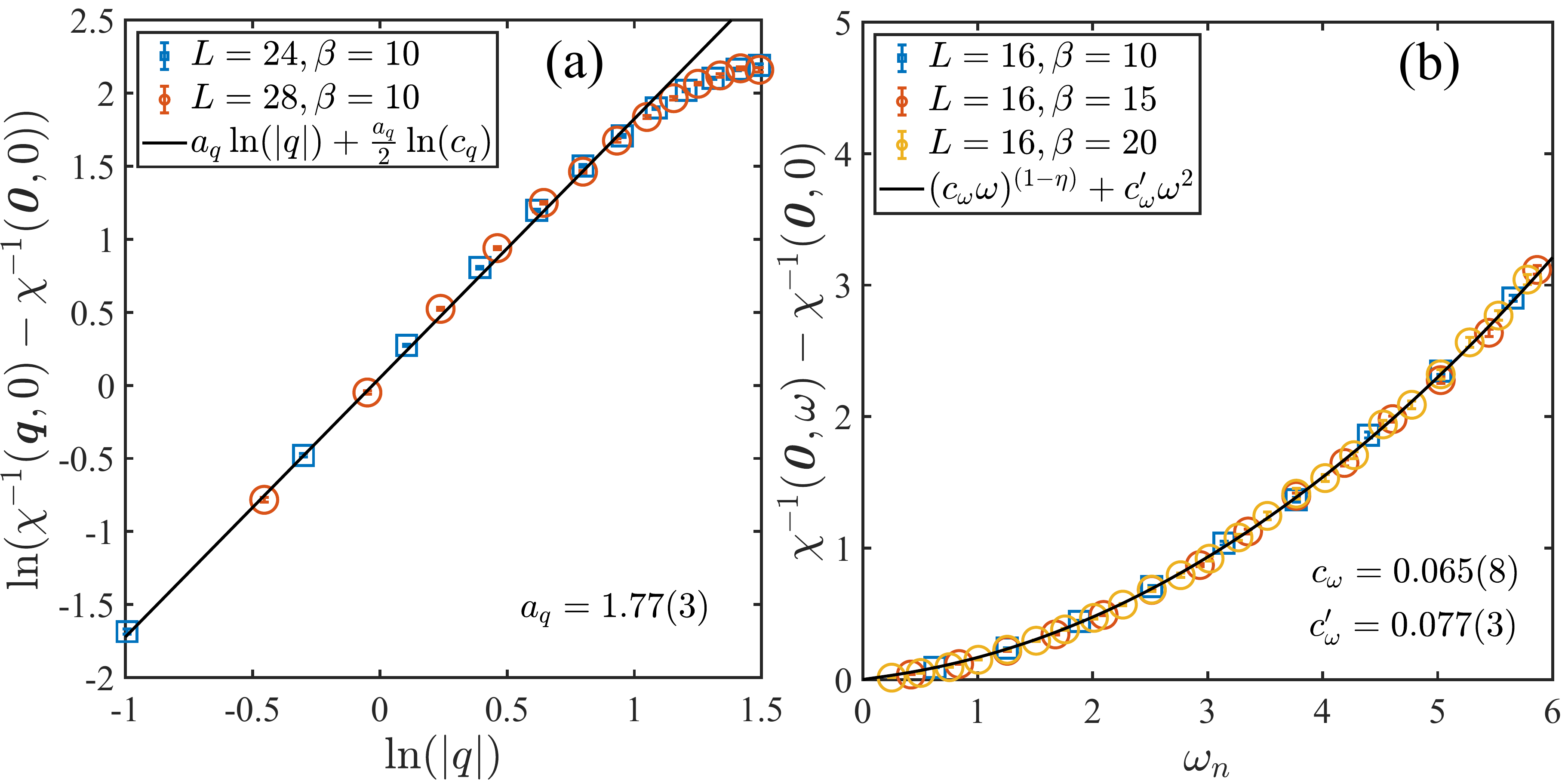}
	\caption{Ising susceptibility data obtained form DQMC at the QCP $h_c=3.32$. (a) $|\mathbf{q}|$ dependence of the bosonic susceptibilities $\chi(T=0,h=h_c,\mathbf{q},\omega=0)$ at the AFM-QCP. The fitting line according to the form in Eq.~\eqref{eq:AFMQCPSQUARE} reveals that there is anomalous dimension in $\chi^{-1}(\mathbf{q})\sim |\mathbf{q}|^{2(1-\eta)}$ with $\eta=0.11(2)$. (b) $\omega$ dependence of the bosonic susceptibilies $\chi(T=0,h=h_c,\mathbf{q}=0,\omega)$ at the AFM-QCP. The fitting line according to the form in Eq.~\eqref{eq:AFMQCPSQUARE} reveals that there is anomalous dimension in $\chi^{-1}(\omega) \sim \omega^{(1-\eta)}$ at small $\omega$ and crossover to $\chi^{-1}(\omega) \sim \omega^2$ at high $\omega$.}
	\label{DQMCFITING}
\end{figure}

In the main text, we have discussed the dynamic Ising spin susceptibility, $\chi(T,h,\mathbf{q},\omega_n)$, and performed the quantum critical scaling analysis by the data obtained from EQMC. And for the sake of saving space, we didn't show the corresponding DQMC results, which we present in this section of SI. The dynamic Ising spin susceptibility obtained from DQMC at quantum critical region $h=h_c=3.32$ are shown in Fig.~\ref{DQMCFITING}, the fitting form of the $\chi$ is still as that in the main text, 
\begin{equation}
\chi(T,h_c,\mathbf{q},\omega_n) \nonumber =\frac{1}{c_{t}T^{a_t}+(c_q |\mathbf{q}|^{2} + c_{\omega}\omega)^{1-\eta}+c'_{\omega}\omega^{2}},
\label{eq:AFMQCPSQUARE}
\end{equation}
with the momentum $|\mathbf{q}|$ is measured with respect to the hot spot $\mathbf{K}$.

\begin{figure*}[htp!]
	\centering
	\includegraphics[width=\textwidth]{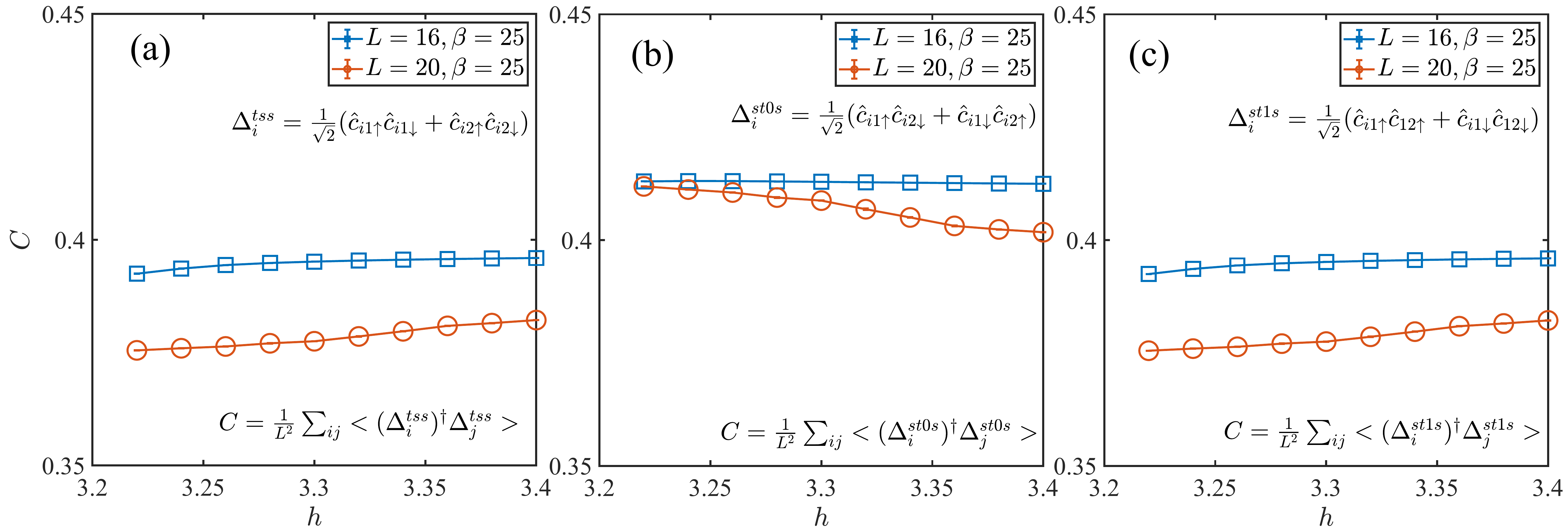}
	\caption{Static pairing-correlation function $C=\frac{1}{L^2}\sum_{ij} \left< \Delta^{\dagger}_i \Delta_j \right >$ for order parameters defined in Eq.~\eqref{scorder} as a function of transverse field $h$. No enhancement of pairing correlation functions is observed.}
	\label{paircor}
\end{figure*}
In Fig.~\ref{DQMCFITING}, We detect the power-laws in momentum dependence and frequency dependence of the bosonic susceptibilities $\chi(T=0,h=h_c,\mathbf{q},\omega=0)$ by fitting  $\chi^{-1}(T,h_c,|\mathbf{q}|,0)-\chi^{-1}(T,h_c,0,0)=c_q |\mathbf{q}|^{2(1-\eta)}$ and $\chi^{-1}(T,h_c,0,\omega)-\chi^{-1}(T,h_c,0,0)=(c_{\omega}\omega)^{1-\eta}+ c'_\omega\omega^{2}$ at QCP located by DQMC. We obtained the values of the coefficients $a_q=2(1-\eta)=1.77(3)$, $c_\omega=0.065(8)$ and $c'_\omega=0.077(3)$. Most importantly, the anomalous dimension $\eta=0.11(2)$ is obtained from DQMC data. These results are consistent with that obtained in our EQMC data showed in the main text.

\section{SUPERCONDUCTIVITY}
We now discuss the superconducting properties close to the $2\mathbf{Q}=\Gamma$ AFM QCP. As described in previous FM-QCP study Ref.~\cite{Xu2017}, in such type of two layered band structure of spin fermion model, the obvious superconductivity instabilities usually happen at very strong coupling strength ($\xi/t=6$ for example) and relatively low temperature, and the spin fluctuations could most likely induce an attractive interaction in the on-site s-wave channel and after that  other channels with even weaker strength. 

To probe for superconducting tendencies near the QCP, we calculate three types of pairing correlations $C=\frac{1}{L^2}\sum_{ij} \left< \Delta^{\dagger}_i \Delta_j \right >$ for on-site s-wave channels, and the superconducting order parameters $\Delta$ is given by
\begin{align}
&\Delta_i^{tss}=\frac{1}{\sqrt{2}}\left(\hat{c}_{1\uparrow}\hat{c}_{1\downarrow}+\hat{c}_{2\uparrow}\hat{c}_{2\downarrow}\right ), \nonumber \\
&\Delta_i^{st0s}=\frac{1}{\sqrt{2}}\left(\hat{c}_{1\uparrow}\hat{c}_{2\downarrow}+\hat{c}_{1\downarrow}\hat{c}_{2\uparrow}\right ), \nonumber \\
&\Delta_i^{st1s}=\frac{1}{\sqrt{2}}\left(\hat{c}_{1\uparrow}\hat{c}_{2\uparrow}+\hat{c}_{1\downarrow}\hat{c}_{2\downarrow}\right ),  \label{scorder}
\end{align} 
where $\Delta_i^{tss}$ is orbital-triplet, spin-singlet channel s-wave order parameter, $\Delta_i^{st0s}$ is orbital-singlet, spin-triplet channel s-wave order parameter and $\Delta_i^{st1s}$ is orbital-singlet, spin-triplet channel s-wave order parameter, on lattice site $i$ and $1$ and $2$ are the layer indices. In Fig.~\ref{paircor}, we show of these three types of static pairing-correlation function $C$ obtained from DQMC data at spin fermion coupling strength $\xi=1.0$ and $\beta=25$ across the AFM-QCP $h_c=3.32$ determined in DQMC simulation.

The pairing-correlation function $C$ in Fig.~\ref{paircor} shows that the pairing correlations do not grow with the system size, indicating that the superconducting instabilities are still weak with our current parameter set. Of course, there is high possibility, as we have seen in the FM-QCP case in Ref.~\cite{Xu2017}, that enhancing the strength of spin-fermion coupling $\xi/t$ much larger than 1 would enhance the superconducting instabilities, but the purpose of this paper is to have a pristine QCP region such that the quantum critical scaling behaviors can be determined within accessible parameter sets, and we found that our current choice of parameter served the purpose.

\end{document}